\documentclass[
reprint,
 groupedaddress,
 amsmath,amssymb,
 aps,
 prd,
 floatfix
]{revtex4-2}
\usepackage{booktabs}
\usepackage{siunitx}
\usepackage{array}
\usepackage{makecell}
\newcolumntype{L}[1]{>{\raggedright\arraybackslash}p{#1}}
\usepackage[utf8]{inputenc} 
\usepackage[T1]{fontenc}
\usepackage{graphicx}
\usepackage{amsmath,amssymb}
\usepackage{xcolor}
\usepackage{hyperref}
\usepackage{natbib}
\usepackage{orcidlink}

\usepackage{graphicx}
\usepackage{dcolumn}
\usepackage{bm}
\usepackage{listings}

\usepackage{array}
\usepackage{booktabs}
\usepackage{siunitx}

\newcommand{{\araa}}{Annual Review of Astronomy and Astrophysics}
\newcommand{{\mnras}}{Monthly Notices of the Royal Astronomical Society}
\newcommand{{\aap}}{Astronomy \& Astrophysics}
\newcommand{{\apjl}}{Astrophys. J. Lett.}
\newcommand{{\apjs}}{Astrophys. J. Sup.}
\newcommand{{\aaps}}{Astron. Astrophys. Suppl. Ser.}
\newcommand{\new}[1]{{#1}}
\begin{document}

\title{The Impact of Errors in the Shape Function of Rotating Neutron Stars in the Oblate Schwarzschild Approximation}

\author{John Ming Ngo \orcidlink{0009-0004-5558-248X}}
 \email{mjngo@ualberta.ca}%

\author{Charlee Amason \orcidlink{0009-0008-7788-2992}}%

\author{Sharon M. Morsink\orcidlink{0000-0003-4357-0575}}%
 \email{morsink@ualberta.ca}
 \affiliation{Department of Physics, University of Alberta, Edmonton, Alberta, Canada}

\date{\today}

\begin{abstract}

Measurements of thermal X-ray flux emitted by neutron stars can constrain their radii and the equation of state of cold dense matter. Accurate flux modeling of rapidly rotating neutron stars requires an approximation for their oblate surface shape, usually given by a shape function. We examine the accuracy of five shape functions by comparing them against numerical relativistic stellar surfaces for seven representative equations of state over a broad range of compactness and spin. We characterize their errors in the surface radius and their derivative with respect to colatitude, test the extent of quasi-universality, and propagate these errors into the differential solid angle $d{\Omega}$, a geometric factor underlying the observed flux, under the oblate Schwarzschild approximation. We find that the shape functions produce distinct error patterns associated with their forms and parameter ranges. These errors remain structured in compactness, spin, and colatitude. Simple polynomial corrections substantially reduce the radius errors of several shape functions, with most corrected errors within approximately $0.5\%$. We show that errors in $d{\Omega}$ may be understood as coming from an area contribution, which scales with twice the relative radius error, and a projection contribution controlled primarily by the error in the surface derivative. The projection contribution grows towards the limb, where small surface errors can produce much larger local errors in $d{\Omega}$. We provide a set of guidelines for the construction of improved quasi-universal shape functions that can be used in future radius estimation efforts. Our recommendations include that both the shape and its derivative be modeled accurately, that more accurate predictions of the polar radius be developed, and that the shape function be calibrated over the same compactness-spin range that is to be tested.

\end{abstract}

\maketitle

\section{Introduction}

The cores of neutron stars are both cold and dense, with densities that are many times larger than nuclear saturation, and thermal energies much lower than the Fermi energy. This range of physical conditions is not well constrained by nuclear theory or laboratory experiments.  A large range of equations of state (EOS) for cold, dense matter have been proposed \cite{COMPOSE_Oertel:2016bki}. Each proposed EOS is mapped to a different curve of possible neutron star masses and radii \cite{2012ARNPS..62..485Lattimer} using the equations of relativistic stellar structure. As a result, observations of the masses and radii of many neutron stars can be used to constrain the EOS of cold, dense matter \cite{1992Lindblom,2012Lindblom}. 

Observations of thermal X-rays emitted from the surfaces of neutron stars, either pulsed or unpulsed, provide a method for estimating the masses and radii of neutron stars and the nuclear EOS. The X-rays travel from the surface of the neutron star to the telescope on paths that are curved by the star's gravitational field.  The gravitational bending of the light is related to the compactness ratio of the neutron star, $M/R_{\rm eq}$, where $M$ is the mass, and $R_{\rm eq}$ is the equatorial radius. (We use units where $G=c=1$.) Doppler effects scale as $v \sim 2 \pi \nu R_{\rm eq}$, where $\nu$ is the star's spin frequency. These relativistic effects can be used to constrain the neutron star's mass and equatorial radius \cite{2016Watts}. In a few cases such as with binary systems, observations of the radio pulsations provide an independent, precise measurement of the mass \cite{2016ARA&A..54..401Ozel}, allowing for stronger constraints on the radius.

Observations of unpulsed X-ray emission from the  
 entire surface of the neutron star by X-ray telescopes such as Chandra and XMM-Newton can be used to determine the flux and effective temperature of the emission, given an atmosphere model \cite{OBS_HEINKE_QLMXB_2006ApJ...644.1090H}. Combining this spectral data with an accurate distance leads to an estimate of the luminosity radius, $R_L = R/\sqrt{1-2M/R}$, which leads to fairly broad constraints \cite{2018Steiner} on the EOS due to the lack of an independent mass measurement. Extra uncertainties are introduced if the neutron star is rotating \cite{LC_2015_Baubock_Thermal} and/or has undetected surface inhomogeneities \cite{2016Elshamouty}, which could lead to biased inferences of the radius. A recent analysis \cite{2026Kazantsev} of Chandra observations of the quiescent low-mass X-ray binary neutron star 47~Tuc~X7 took into account the unknown rotation of this star and possible surface anisotropies and found a large range of radii corresponding to a precision of approximately 20\%.
 In the case of X-ray burst sources,  it has been shown \cite{2020A&A...639A..33Suleimanov,2017A&A...608A..31Nattila} that the inclusion of rotation introduces a large uncertainty to the EOS constraints. It is expected that NewAthena (with a launch date in the 2030s) will have a large enough effective area to provide stronger constraints \cite{2025NatAs...9...36Cruise} on the EOS using observations of unpulsed surface X-ray emission from many neutron stars. 
 
Pulse-profile modeling \cite{2016Watts,OS_2019_NICER} is used to infer the masses and radii of rotating X-ray pulsars using spectral-timing data collected by telescopes such as the Neutron Star Interior Composition Explorer (NICER)
\cite{INSTR_NICER_2016SPIE.9905E..1HG}. NICER observations have been used to provide estimates of the masses and radii of a few rotation-powered pulsars, including PSR~J0030+0451 \cite{2019ApJ...887L..24Miller,2019ApJ...887L..21Riley,2026Kini}, PSR~J0740+6620 \cite{2021ApJ...918L..28M,2021ApJ...918L..27Riley,2024ApJ...974..295Dittmann,2024ApJ...974..294Salmi}, PSR~J1231-1411 \cite{2024ApJ...976...58Salmi}, PSR~J0437-4715 \cite{2024ApJ...971L..20Choudhury,2026ApJ..1000L..48Miller},
PSR~J0614-3329 \new{\cite{2025ApJ...995...60Mauviard, 2026arXiv260900965M}, and PSR~J1614-2230 \cite{2026arXiv260900172M}}. These pulsars have moderate spin frequencies that range from 174 to 347 Hz. Planned observatories, such as the enhanced X-ray Timing and Polarimetry (eXTP) telescope \cite{2025SCPMA..6819503Li}, will be launched in the 2030s and will extend these types of measurements to more neutron stars, including the more rapidly rotating (300 - 600 Hz) accretion-powered millisecond-period X-ray pulsars (AMXPs).

Since many neutron stars that emit X-rays from their surface are rotating,  it is important to investigate how rotation affects the computation of the detected flux. In particular, the photon geodesics will be altered by the gravitational field of a rotating object, compared to the photon geodesics computed in the spherically symmetric Schwarzschild metric appropriate for a non-rotating neutron star. Early work \cite{1989ApJ...339..279Chen,2000ApJ...531..447Braje} modeled these effects by embedding a spherical surface in the metric of a Kerr black hole. However, the surface of a rotating neutron star is not spherically symmetric, and the Kerr metric is only a good description asymptotically far from a rotating neutron star. A numerical computation making use of the correct oblate shape and metric arising from the relativistic stellar structure equations \cite{BK_2007_Cadeau_Raytracing} showed that the most important effect on pulse profiles is the oblate shape of the rotating star, and that the choice of metric only adds small corrections. This motivated the construction of the oblate Schwarzschild (OS) approximation \cite{OS_SF_2007_Morsink}, where an oblate surface is embedded in the Schwarzschild metric, and Doppler effects are added using the same methods introduced in the Schwarzschild plus Doppler (SD) \cite{1998ApJ...499L..37Miller,BK_Poutanen_Doppler_2003MNRAS.343.1301P} approximation. Although the deviations from a spherical shape are small, the initial conditions for angles at which photons can be emitted with respect to the surface are quite sensitive to the shape. In the case of the neutron stars with moderate spins near 200 Hz observed by NICER, it has been shown \cite{OS_2019_NICER} that the OS approximation should be used in pulse-profile modeling of NICER data. Further calculations \cite{LC_2018_Pihajoki_ARCMANCER_RT,LC_2018_Nattila_RT} for more rapidly rotating neutron stars with spin frequencies faster than 600 Hz have shown that the OS formalism continues to provide an accurate approximation. However, another calculation of geodesics in spacetimes with differing quadrupole moments \cite{2021MNRAS.505.2870Oliva} has illustrated that small errors are introduced by using the Schwarzschild metric, so this merits a more detailed analysis elsewhere.

The OS approximation is computationally inexpensive because the calculation of geodesics in the Schwarzschild metric is possible through a simple integral \cite{BK_PECHENICK_HOTSPOTS_1983ApJ...274..846P}. In this approximation, the oblate shape is described by the stellar surface, expressed as a radius $R(\theta)$ as a function of colatitude. This surface can be computed directly with a relativistic rotating-stellar-structure code, but doing so requires choosing a particular EOS. Shape functions are introduced to avoid making the surface calculation EOS-specific. They replace the tabulated surface with an approximate relation for the stellar shape dependent on two dimensionless parameters: the compactness, $\zeta$,
\begin{equation}
    \zeta = \frac{M}{R_{\rm eq}}
    \label{eq:zeta}
\end{equation}
and the spin parameter, $\epsilon$, defined by
\begin{equation}
 \epsilon = \frac{\Omega^2 R_{\rm eq}^3}{M}.
 \label{eq:epsilon}
\end{equation}

The usefulness of this approach depends on an assumption of quasi-universality: after the surface is expressed in terms of these dimensionless quantities, the remaining dependence on the EOS should be minimal. If this assumption were exact, different shape functions fitted to different EOS samples would reproduce the same surface for the same $(\theta,\zeta,\epsilon)$. In practice, several shape functions have been introduced \new{\cite{OS_SF_2007_Morsink, SF_2013_Baubock,SF_2014_AlGendy,SF_2021_Silva,SF_Papigkiotis_ML, 2026arXiv260900172M}}, and they make different assumptions about the colatitude dependence, the dimensionless-parameter dependence, and the EOS set used for calibration. These differences lead to disagreements between their predicted surfaces. For any particular EOS, there may also be a deviation from the quasi-universal shape captured by a given fit.

Until now, there has never been a detailed analysis of how accurate a shape function needs to be in order to provide an accurate prediction of the observed flux. The purpose of this paper is to provide a careful study of the systematic errors in the observed flux introduced by errors in the shape function. If a small surface element, $dS$, on the star emits light, the observed flux, $dF$, can be written schematically as 
\begin{equation}
    dF = \Upsilon d\Omega,
\end{equation}
where $d\Omega$ is the solid angle subtended by the surface, and $\Upsilon$ is a function that includes the specific intensity of the atmosphere, redshift, Doppler, and Lorentz factors, and the light travel times of photons emitted from different parts of the star. Errors in the oblate shape of the star induce errors in the observed solid angle of an area element, leading to a fractional error in the observed flux, $\Delta(dF)/dF$,
\begin{equation}
    \frac{\Delta(dF)}{dF} = \frac{\Delta (d\Omega)}{d\Omega} + \frac{\Delta(\Upsilon)}{\Upsilon},
\end{equation}
where $\Delta$ is the error in a quantity. This paper focuses on a detailed computation of $\Delta(d\Omega)$, the error in the solid angle caused by errors in the shape function, assuming the OS approximation. \new{We focus on local errors for infinitesimal surface elements. For a more realistic extended emitting region, such as a hot spot, the overall error in the total flux depends on an integration over the surface elements weighted by their contribution to the total flux. Many large-error regions shown in this paper come from regions with small $d\Omega$, such as the limb; thus, such an integration will have smaller errors than for infinitesimal spots.}

We note that there are other known systematic errors present in the OS approximation, aside from the errors in the shape function, which are encompassed in the $\Delta(\Upsilon)/\Upsilon$ term. Previous work \cite{2025ApJ...994..163Jakab} has shown that there are systematic errors introduced in the original \cite{OS_SF_2007_Morsink} treatment of the redshift in the OS approximation, and provides a simple universal correction factor. A more accurate approximation \cite{SF_Papigkiotis_ML} for the variation of the acceleration due to gravity on the rotating star's surface has been computed. The surface gravity is a parameter required in the atmosphere models, such as in Hydrogen atmospheres \cite{OBS_HEINKE_QLMXB_2006ApJ...644.1090H, 2001MNRAS.327.1081Ho}. These Hydrogen atmosphere models also have inherent assumptions (such as deep heating) that could lead to systematic errors \cite{2019ApJ...872..162Baubock,2020A&A...641A..15Salmi,2025ApJ...982..112Zhao}; however, a comparison \cite{2023ApJ...956..138Salmi} with atmospheres with extra beaming parameters showed that radius inference is not biased by small departures from a standard deeply heated Hydrogen atmosphere.

The relationship between the systematic uncertainties in the flux and the accuracy and precision of the radius (and other parameters) estimated from a Bayesian analysis is complicated. Tests of Bayesian parameter estimation codes 
against synthetic NICER data
(see, for example 
\cite{2013ApJ...776...19Lo,2015ApJ...808...31Miller,2021ApJ...914L..15Bogdanov,2025MNRAS.537.3769Bootsma,2025ApJ...982..112Zhao}) depend on the number of photons assumed to originate from the hot spot and the background. It is beyond the scope of the present work to attempt a Bayesian analysis to understand the impact of the systematic errors introduced by the
shape function. Instead, our goal is to provide an understanding of the magnitude and dependencies of these systematic errors with the goal that they should be smaller than the statistical errors in any data set.

We have a few main goals in this paper which are addressed after introducing the theoretical framework and our methodology in Section~\ref{sec:methods}. In Section~\ref{sec:shape_function_errors} we will examine the sizes and morphologies of the errors present in all of the shape functions that have been proposed
\new{at the time of the submission of this paper}. In Section~\ref{sec:solid_angle_errors} we provide the details of an error propagation analysis to show how errors in the shape function lead to errors in the solid angle, a key factor in the observed flux in the OS approximation. In this section, we explicitly compute the errors in the solid angle caused by shape function errors. In Section~\ref{sec:conclusions} we provide detailed recommendations for how to construct more accurate shape functions appropriate for use with rapidly rotating neutron stars. \new{We also provide a translation between the notation conventions used by different papers for the OS approximation in the appendix.} We will not (in this paper) provide a new, improved shape function. Instead, we provide the tools that any research group can use to construct their own shape function.

\section{Methods}
\label{sec:methods}

In this paper, we compare approximate shape functions against a reference surface computed numerically using a relativistic stellar structure code and a specified EOS. We use the Rotating Neutron Star (\texttt{RNS}) code to compute the reference surfaces. Given an EOS, a central density, and a spin frequency $\nu$, \texttt{RNS} solves the Einstein field equations for an axisymmetric rotating neutron star and returns the gravitational mass $M$, the equatorial radius $R_{\rm eq}$, and the tabulated surface $R_{\rm RNS}(\theta)$, which we take as the reference surface. We then evaluate each shape function from the literature using the same $M$, $R_{\rm eq}$, and spin frequency. This gives a second surface, $R_{\rm sf}(\theta)$, for the same model. The mass, equatorial radius, and spin are held fixed, so the difference between $R_{\rm sf}(\theta)$ and $R_{\rm RNS}(\theta)$ gives us the error introduced by using the shape function in place of the \texttt{RNS}-computed surface for the given EOS. This allows us to determine how closely each shape function obeys quasi-universality within our parameter space. 

We make this comparison in stages. First, we compare the radii as functions of colatitude by taking their difference as the error. This shows the scale and morphology of the radial error over colatitude. Second, we compare the surface derivatives, since the derivative affects the local surface normal in the OS approximation. Third, we use our computational implementation of the OS approximation, Neutron Star Schwarzschild With Oblate Rotational Deformations (NS-SWORD)\footnote{\url{https://github.com/John-Ming-Ngo/NS-SWORD}}, to compute the differential solid angle $d\Omega$ for both the \texttt{RNS} surface and the shape function surface and take their difference as the error. We note that with increasing compactness, more of the back of the star is visible to an observer. At $\zeta \leq 1/3.52$, the entire back of the star is visible to an observer, and past that point, many points on the back of the star are visible from two separate directions to an observer, a phenomenon known as double imaging. For the solid-angle calculations with NS-SWORD, this has not been implemented, so we restricted the models to $\zeta \leq 1/3.52$.  This lets us compute how differences in $R(\theta)$ and its derivative propagate into the solid angle used in the flux calculation.

The rest of this section goes into greater depth about the ingredients used in these comparisons. In the following subsections, we first review the OS geometry and the differential solid angle used in NS-SWORD. We then describe how the reference surfaces are computed with \texttt{RNS}. Then, we define the dimensionless compactness and spin parameters used throughout the analysis. We next summarize the approximate shape functions drawn from the literature and the EOS sample used to test them. 


\subsection{Oblate Schwarzschild Approximation}
\label{sec:OSapprox}



\begin{figure}[!ht]
    \centering
    \includegraphics[width=0.75\linewidth]{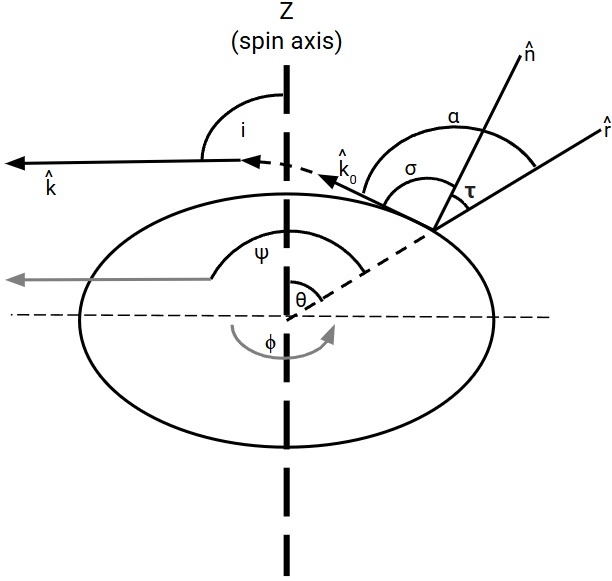}
    \caption{Diagram illustrating the oblate Schwarzschild (OS) geometry. A photon leaves the neutron star surface along $\hat{k}_0$ and reaches the distant observer along $\hat{k}$ after bending by an angle 
    \new{$\psi-\alpha$}, where $\cos\psi = \hat{k}\cdot\hat{r}$. At the emission point, $\hat{r}$ is the local radial direction and $\hat{n}$ is the local surface normal. The angles $\alpha$, $\sigma$, $\tau$, $i$, $\theta$, and $\phi$ are defined in the text. This figure illustrates the notation used throughout the paper. Note that the observer has been placed in the equatorial plane ($i=90^\circ$) to simplify the drawing.}
    \label{fig:OS_geometry}
\end{figure}

The OS approximation, as illustrated by Figure \ref{fig:OS_geometry}, starts with the assumption that the Schwarzschild metric describes the exterior region of a rotating neutron star \cite{OS_SF_2007_Morsink}. This is motivated by studies of the exact numerical metric \cite{BK_2007_Cadeau_Raytracing}. The coordinate $\theta$ is the colatitude, measured from the north spin axis so that the equator is at $\theta=\pi/2$, and $\phi$ is the azimuthal angle measured around the spin axis. The surface of the star is defined by values of $r = R(\theta)$, where $r$ is the radial coordinate used in the Schwarzschild metric and $R(\theta)$ is the shape function that defines the surface of the rotating star. The numerical computation of the surface $R(\theta)$ is reviewed in Section \ref{sec:compmethod} and the approximate shape functions are reviewed in Section \ref{sec:shapes}.

As in the spherical Schwarzschild approximation \cite{BK_Poutanen_Doppler_2003MNRAS.343.1301P}, a photon is emitted from an infinitesimal hot spot with coordinates $\theta$ and $\phi$ on the surface of the star and detected by the observer located at colatitude $i$ (the inclination angle) and azimuthal angle $\phi=0$. Spatial direction vectors $\hat{r}$ and $\hat{k}$ point in the local radial direction of the spot and the observer, respectively.

 The photon is emitted into an initial spatial direction $\hat{k}_0$, at an angle $\alpha$ from the local radial direction. The local azimuthal angle $\lambda$, defined around the radial direction, is defined so that $\lambda=0$ corresponds to a photon that is initially directed towards the north spin axis.

The gravitational field bends the photon's direction through an angle $\psi-\alpha$, where the bending angle $\psi$ is defined through spherical trigonometry
\begin{equation}
    \cos \psi = \cos i \cos \theta + \sin i \sin\theta \cos\phi.
    \label{eq:psi}
\end{equation}
The local azimuthal angle $\lambda$ is defined through the spherical triangle relation
\begin{equation}
    \cos i = \cos \theta \cos \psi + \sin \theta \sin \psi \cos \lambda,
    \label{eq:lambda}
\end{equation}
which follows from the coplanarity of the vectors $\hat{r}$, $\hat{k}$, and $\hat{k}_0$ in the Schwarzschild metric. In our examinations of the errors introduced by the OS approximation, we consider the values of $\theta$, $\phi$, and $i$ to be known exactly, so that $\psi$ and $\lambda$ have no error.

Errors in the shape function introduce errors in the angle $\alpha$, since it is defined implicitly through the integral relation
\begin{eqnarray}
    \psi(b,R(\theta)) = b \int_{R(\theta)}^\infty \frac{dr}{r^2}
    \left[1-\frac{b^2}{r^2}\left(1-\frac{2M}{r}\right)\right]^{-1/2},
    \label{eq:psi_integral}
\end{eqnarray}
and the impact parameter $b$ is defined by
\begin{eqnarray}
    \sin\alpha = \frac{b}{R(\theta)}
    \sqrt{1-\frac{2M}{R(\theta)}}.
    \label{eq:alpha}
\end{eqnarray}
These equations describe the initially outgoing photon geodesics in the Schwarzschild metric.
 The full OS approximation also requires an examination of initially ingoing geodesics, which is discussed in more detail elsewhere \cite{OS_SF_2007_Morsink,OS_2019_NICER} and will not be repeated here.

The OS approximation requires the introduction of two more angles, $\sigma$ and $\tau$ (using the notation of \cite{OS_2019_NICER}), describing the oblate surface. The angle $\sigma$ is the angle between the initial photon direction $\hat{k}_0$ and the local normal to the surface, $\hat{n}$. The angle between the local radial and normal direction vectors is $\tau$. \new{Since different papers use different symbols for these angles, Table~\ref{tab:notation_translation} in the appendix provides a translation between our notation and the conventions used in those papers.} These angles are related by
\begin{equation}
\cos\sigma
=
\cos\alpha\cos\tau
+
\sin\alpha\sin\tau\cos\lambda ,
\label{eq:cossigma_alpha_tau_lambda}
\end{equation}
and
\begin{equation}
\cos\tau
=
(1+ q^2(\theta))^{-1/2},
\label{eq:tau}
\end{equation}
where the slope factor $q(\theta)$ is defined by
\begin{equation}
    q(\theta) = \left(1 - \frac{2M}{R(\theta)}\right)^{-1/2} \frac{R'(\theta)}{R(\theta)} ,
    \label{eq:surface_slope_q}
\end{equation}
where $R'(\theta)$ is the derivative of $R(\theta)$ with respect to $\theta$,
and
$q(\theta)$ is positive in the northern hemisphere and negative in the southern hemisphere.
As a result, errors in $R(\theta)$ and $R'(\theta)$ lead to errors in the angles $\tau$ and $\sigma$.

In the OS approximation, the observed flux of photons with energy $E$ emitted from a region on the star with angular extent $d\theta$ and $d\phi$ is
\begin{equation}
    dF_E  = \frac{\Gamma \mathcal{D}}{(1+z)^3} I'(E',\sigma') d\Omega,
\end{equation}
where the energy of the emitted photon is $E' = (1+z) E$ \cite{2025ApJ...994..163Jakab}, $\Gamma$ and $\mathcal{D}$ are the Lorentz and Doppler factors \cite{OS_2019_NICER}, $I'(E',\sigma')$ is the specific intensity in the local inertial frame momentarily co-moving with the surface of the star, and the differential solid angle $d\Omega$ is defined by \cite{OS_SF_2007_Morsink,OS_2019_NICER}
\begin{equation}
d\Omega =
\frac{R^2(\theta)\sin\theta\,d\theta\,d\phi}{\cos\tau \, D^2}
\times
\cos\sigma
\times
\frac{1}{(1-\frac{2M}{R(\theta)})} \left|  \frac{\partial \cos\alpha}{\partial \cos\psi} \right|_R.
\label{eq:domega_split}
\end{equation}
We split the differential solid angle in equation (\ref{eq:domega_split}) into three terms. The first term is the surface area of a differential area element divided by the square of the distance $D$ to the star, where the $\cos\tau$ term is due to the area element of an oblate surface being larger than that of a spherical surface. The second term, $\cos\sigma$, is the oblate projection factor, so that light emitted normal to the surface leads to a larger subtended solid angle. The third term encompasses the gravitational self-lensing of light emitted at $R(\theta)$ in the Schwarzschild spacetime.

In this paper, we compute the errors in the differential solid angle due to errors in the shape function and its derivatives. The prefactor $\Upsilon = \frac{\Gamma \mathcal{D}}{(1+z)^3} I'(E',\sigma')$ also has errors introduced by errors in the shape function. These errors include errors in the redshift term, which has been examined carefully elsewhere \cite{2025ApJ...994..163Jakab}. We defer a study of errors in the special relativistic prefactors and the beaming pattern of the specific intensity to future work. The specific form of the error in the differential solid angle will be derived in Section \ref{subsec:analytic_differential_solid_angle}.

\subsection{Computation of the Shape of a Rotating Star}
\label{sec:compmethod}

The shape of a rotating star can be computed by numerically calculating the relativistic equations of stellar structure for a rigidly rotating neutron star, given an EOS. The metric for a stationary, axisymmetric spacetime is given by \cite{1971ApJ...167..359Bardeen}
\begin{eqnarray}
    ds^2 &=& - e^{2\Phi} dt^2 + \bar{r}^2 \sin^{2}\theta B^2 e^{-2\Phi}\left( d\phi - \omega dt\right)^2 \nonumber\\
    &&+ e^{2\eta -2\Phi}\left( d\bar{r}^2 + \bar{r}^2 d\theta^2\right),
    \label{eq:metric}
\end{eqnarray}
where the 4 metric functions $\Phi$, $B$, $\omega$, and $\eta$ depend only on the coordinates $\bar{r}$ and $\theta$. In the limit of zero rotation, this metric approaches the isotropic Schwarzschild metric, and $\bar{r}$ approaches the isotropic radial coordinate \cite{1973grav.book.....Misner}.
The circumferential radius $r$ (with the property that a circle with radius $r$ has a circumference of $2 \pi r$) used in the OS approximation is defined by $r = \bar{r} B e^{-\Phi}$. Far from the star, the metric potential $\Phi$ approaches the Newtonian gravitational potential. The metric potential $\omega$ encodes the dragging of inertial frames. 

Given an EOS, the Einstein field equations 
for the metric (\ref{eq:metric})
can be solved numerically using one of many computer codes such as 
{\texttt{Lorene}} \cite{1998PhRvD..58j4020Bonazzola} or {\texttt{RNS}} \cite{RNS_Implementation_Stergioulas_1995ApJ...444..306S}, which have been numerically compared with each other \cite{1998A&AS..132..431Nozawa} to verify their accuracy. We make use of an implementation of {\texttt{RNS}} \footnote{\url{https://github.com/rns-alberta/rns}}. 

The {\texttt{RNS}} code is based on a Green's function method  \cite{RNS_1989_KEH,RNS_1994_CST} and described in more detail elsewhere \cite{SF_Papigkiotis_ML}. Since we aim to model old, cold neutron stars, we restrict our computations to rigidly rotating neutron stars. The oblate surface of the star is found by solving for the surface of zero specific enthalpy. The details of the definition of the specific enthalpy and the methods used to solve for the surface are the same as detailed in the paper \cite{SF_Papigkiotis_ML}. In particular, first the equation for the surface in the $\bar{r}$ coordinate is found, and then it is transformed to the form $R(\theta)$, which gives the value of the circumferential radial coordinate at each value of colatitude on the surface of the rotating star.

The {\texttt{RNS}} code computes the star's properties using a grid based on the cosine of the colatitude, $\mu = \cos\theta$, and a compactified radial coordinate, $s$ \cite{RNS_1994_CST}. Both coordinates $\mu$ and $s$ range from 0 to 1, and are divided into \texttt{MDIV} $\mu$ divisions and \texttt{SDIV} $s$ divisions. We compiled {\texttt{RNS}} with two separate resolutions for different sections of the analysis. For analyzing the errors in radius,
derivative of the radius, and the differential solid angle, we utilized a computational grid resolution of \texttt{MDIV}$\times$\texttt{SDIV} $=261\times2081$. A larger \texttt{SDIV} was necessary to minimize noise in the derivative of the radius below the order of magnitude of the errors shown in this paper. For the universality analysis, which only deals with the radial errors, we made use of a \texttt{MDIV}$\times$\texttt{SDIV} $=261\times521$ for speed and simplicity, as noise in the derivative of the radius does not matter in that context.


\subsection{Dimensionless Parameters}
\label{sec:parameters}

The goal of an approximate shape function is to provide a simple model of the $\theta$-dependence of the oblate shape of the neutron star that does not introduce any new parameters to the Bayesian parameter estimation process. Since the spin frequency, $\nu$, of a pulsar is normally known to high precision, while the mass $M$ and equatorial radius $R_{\rm eq}$ are already free parameters, it makes sense to use the two dimensionless parameters $\zeta$ and $\epsilon$ defined in equations (\ref{eq:zeta}) and (\ref{eq:epsilon}).
\new{However, even when the spin frequency is known, $\zeta$ and $\epsilon$ vary with the unknown $M$ and $R_{\rm eq}$. In the case of neutron stars with no pulsations, the spin frequency is not known, which adds an extra free parameter in determining $\epsilon$.}
The compactness controls the magnitude of the gravitational lensing of photons emitted at the equator. The spin parameter is the ratio of the centrifugal to the gravitational force at the equator in the limit of Newtonian gravity. 

The range for the values of $\zeta$ and $\epsilon$ can be understood by examining the spin and mass measurements for a sample of representative neutron stars, shown in Table \ref{tab:obs_ns_examples}. In Table \ref{tab:obs_ns_examples}, we list the spin frequency and mass (where available). The observational classifications are abbreviated as: accreting millisecond X-ray pulsar (AMXP), thermonuclear X-ray burst sources (Bursts), black widow or redback (Spider), quiescent low-mass X-ray binary (qLMXB), radio pulsar (Radio), and rotation-powered pulsar (RPP). We include the lowest known mass neutron star, PSR J0453+1559B (assuming that it is not a white dwarf) \cite{EX_NS_SMALL_2015ApJ...812..143M}. That mass was determined through observations of its pulsar companion. The neutron star with the highest mass determined through radio timing of Shapiro delay \cite{OBS_MASS_SHAPIRO_Cromartie_2019,EX_NS_LARGE_2021ApJ...915L..12F} is PSR J0740+6620. The pulsar PSR J0952-0607 has a higher mass (and spin), but the mass is determined through a more model-dependent method \cite{2022ApJ...934L..17Romani}.

\begin{table*}[ht!]
  \centering
  \caption{Representative neutron stars spanning the observed range of mass and spin frequency. 
  The symbol "-" indicates that there is no measured value. 
  One-sigma error bars are shown for the masses.
  The spin frequencies of the non-accreting stars are typically known to much higher precision than shown in this table. 
  }
  \label{tab:obs_ns_examples}
  \begin{tabular}{@{} l l l S[table-format=3.2] l @{}}
    \hline
    Type & Name & $M$ ($M_{\odot}$) & {Spin (Hz)} & Reference \\
\hline
    No Pulsations        & PSR J0453+1559B  & $1.174 \pm 0.004$ & {-}   & \citep{EX_NS_SMALL_2015ApJ...812..143M} \\
    AMXP, Bursts         & IGR J17480-2446  & -           & 11.0     & \citep{2012MNRAS.423.1178Papitto} \\
    RPP, Radio & PSR J0437-4715   & $1.418 \pm 0.044$   & 173.7    & \citep{Ex_NS_NICER_1_174_2024ApJ...971L..18R} \\
    RPP, Radio & PSR J0614-3329   & $1.44 \pm 0.07$          & 317.6    & \citep{Ex_NS_NICER_2_318_2011ApJ...727L..16R,2025MNRAS.536.1467Miles,2025ApJ...995...60Mauviard} 
    \\
    RPP, Radio      & PSR J0740+6620   & $2.08 \pm 0.07$    & 346.5  & \citep{EX_NS_LARGE_2021ApJ...915L..12F} \\
    AMXP, Bursts, qLMXB & SAX J1808-3658  & -           & 401.0    & \citep{EX_NS_XRAYB_2_401_1998Natur.394..344W} \\
    AMXP, Bursts, qLMXB & Aql X-1  & -          & 550.3    & \citep{EX_NS_QLMXB_1_550_2008ApJ...674L..41C} \\
    Bursts & 4U 1730-22      & -          & 584.7 & \citep{EX_NS_XRAYB_1_584_2022ApJ...935..123L} \\
    AMXP, Radio, Spider         & PSR J1023+0038   & $1.65^{+0.19}_{-0.16}$    & 592.4    & \citep{Ex_NS_AMXP_2_592_2019ApJ...872...42S} \\
    Radio, Spider & PSR J0952-0607 & $2.35 \pm 0.11$ & 707.3 &
    \citep{2017ApJ...846L..20Bassa,2026ApJ...996..101R}
    \\
    Radio         & PSR J1748-2446ad & -           & 716.3    & \citep{EX_NS_FASTEST_716_2006Sci...311.1901H} \\
\hline
  \end{tabular}
\end{table*}

The RPPs listed in Table \ref{tab:obs_ns_examples} are three of the pulsars whose masses and radii have been estimated using NICER data and include the slowest and fastest spinning RPPs observed by NICER. The two fastest neutron stars listed (with $\nu > 700$ Hz) are not candidates for a radius estimation analysis; however, we include them to show the range of spin frequencies known at present.

Based on the known values of spin and mass, appropriate ranges of the dimensionless parameters $\zeta$ and $\epsilon$ can be calculated, given a range of possible equatorial radii. A few illustrative values of these parameters are listed in Table~\ref{tab:parameter_space_examples_v2}. The entries for $\nu=200$ Hz and $M = 1.4 M_\odot$ show a typical range of the dimensionless parameters suitable for modeling PSR J0437-4715, for example.

\begin{table}[ht!]
    \centering
    \caption{Illustrative values of neutron star compactness $\zeta$ and spin parameter $\epsilon$ for typical combinations of mass, spin frequency, and radius. This table highlights the range of parameters appropriate for the slowly and rapidly rotating neutron stars discussed in this paper.}
    \label{tab:parameter_space_examples_v2}

    \begin{tabular}{ccccc}
        \hline
$\nu$ (Hz) & $M (M_\odot)$ & $R_{\rm eq}$ (km)& $ \qquad \zeta \qquad $ & $\epsilon$ \\
\hline
200 & 1.0 & 9 & 0.164 & 8.674e-03 \\ 
200 & 1.0 & 12 & 0.123 & 2.056e-02 \\ 
200 & 1.0 & 15 & 0.098 & 4.016e-02 \\ 
200 & 1.4 & 9 & 0.230 & 6.196e-03 \\ 
200 & 1.4 & 12 & 0.172 & 1.469e-02 \\ 
200 & 1.4 & 15 & 0.138 & 2.869e-02 \\
200 & 1.8 & 9 & 0.295 & 4.819e-03 \\ 
200 & 1.8 & 12 & 0.221 & 1.142e-02 \\ 
200 & 1.8 & 15 & 0.177 & 2.231e-02 \\ 
\hline
600 & 1.0 & 9 & 0.164 & 7.807e-02 \\ 
600 & 1.0 & 12 & 0.123 & 1.851e-01 \\ 
600 & 1.0 & 15 & 0.098 & 3.614e-01 \\
600 & 1.4 & 9 & 0.230 & 5.576e-02 \\ 
600 & 1.4 & 12 & 0.172 & 1.322e-01 \\ 
600 & 1.4 & 15 & 0.138 & 2.582e-01 \\ 
600 & 1.8 & 9 & 0.295 & 4.337e-02 \\ 
600 & 1.8 & 12 & 0.221 & 1.028e-01 \\ 
600 & 1.8 & 15 & 0.177 & 2.008e-01 \\ 
        \hline
    \end{tabular}
\end{table}

\subsection{Approximate Shape Functions}
\label{sec:shapes}

All of the shape functions drawn from the literature are constructed with a similar methodology. First, a collection of EOSs is chosen. Then the surfaces, $R(\theta)$, are computed for a wide range of masses and spin frequencies using either a rapid-rotation code (such as \texttt{RNS}) or the Hartle-Thorne \cite{1967Hartle,1968Hartle}
slow-rotation approximation. An angular dependence with coefficients that depend on $\zeta$ and $\epsilon$ is assumed, and a fit to the numerically computed surfaces is found. 

The shape functions make use of the assumption of approximate universality.  Universality is the observation that certain properties of neutron stars follow a relationship independent of the specific equation of state chosen. In this context, it is an empirically based assumption that once the mass, equatorial radius, and spin frequency are known, the other rotational properties of the star (such as the oblate shape) depend only on the dimensionless parameters $\epsilon$ and $\zeta$ in a way that is almost independent of the choice of EOS. The empirical evidence comes from the reported accuracy of each shape function in their original works. Despite this, there are multiple different shape functions which do not agree with each other, prompting this analysis. 

The main difference between the different approximate shape functions lies in the assumed angular dependence, the choices of EOS, and the ranges of parameters used. In the following sections, the various approximate shape functions will be described in chronological order. The acronyms for each shape function are formed from the first letters of the authors' last names.

\new{As an aside, after this paper was submitted, a shape function tailored specifically for the known spin frequency, 317 Hz, of PSR~J1614-2230  was introduced \cite{2026arXiv260900172M}. We do not analyze this shape function here, in part because it is not a general shape function meant for a range of spin frequencies.}
    
\subsubsection{MLCB: Morsink, Leahy, Cadeau, \& Braga (2007) \cite{OS_SF_2007_Morsink}}

The first shape function proposed, MLCB, was introduced \cite{OS_SF_2007_Morsink} as part of the OS approximation.  The MLCB shape function models  $R(\theta)$ as a perturbation from spherical symmetry using  Legendre polynomials, $P_{n}(\cos\theta)$. The formula is 
\begin{equation}
    R(\theta)\;=\;R_{\rm eq}\!\left[1
    +a_{0}P_{0}(\cos\theta)
    +a_{2}P_{2}(\cos\theta)
    +a_{4}P_{4}(\cos\theta)\right],
\label{eq:morsink_shape}
\end{equation}
with coefficients $a_{2n}=a_{2n}(\zeta,\epsilon)$ that depend on polynomials of $\zeta$ and $\epsilon$. The parametrization chosen was an ansatz based on similar parametrizations for other neutron star properties found earlier \cite{1994ApJ...424..846Ravenhall, EOS_LATTIMER_PRAKASH_2001ApJ...550..426L}.

The neutron star surfaces were computed using \texttt{RNS}, using a 
set of equations of state representative of the nuclear theory landscape at the time. The EOS library included the EOS catalog (A, B, C, F, G, L, N, O) of \citet{EOS_ARNETT_1977ApJS...33..415A}, APR \cite{EOS_AKMAL_PhysRevC.58.1804}, a suite of hybrid quark stars with mixed quark-hadron phases, \cite{EOS_ALFORD_2005ApJ...629..969A}, and hyperon-rich models \cite{EOS_LACKEY_PhysRevD.73.024021}.

The main issue with the MLCB shape function is that the fitting procedure did not enforce the condition $R_{\rm eq} = R(\pi/2)$. This is a known input to the shape function, so allowing $R(\pi/2)$ to deviate introduces unnecessary errors. Additionally, the use of a wider range of EOSs, masses, radii, and spin frequencies would be an improvement \cite{SF_2021_Silva}. The MLCB shape function was used in the analysis of RXTE data for the AMXPs SAX~J1808-3658 \cite{2008ApJ...672.1119Leahy,2011ApJ...726...56Morsink}, XTE J1814-338 \cite{2009ApJ...691.1235Leahy}, and XTE~J1807-294 \cite{2011ApJ...742...17Leahy} with spin frequencies ranging from 190 - 401 Hz. Any of the inaccuracies introduced by the shape function are dwarfed by the uncertainties in the atmosphere model for these neutron stars. While the MLCB shape function is not used in any parameter estimation codes which analyze NICER data, we include it for historical completeness. The main interest is that this work showed that the shape could be parameterized in a way that is almost independent of the chosen EOS, making it possible to include the oblate shape without adding extra parameters to a parameter estimation code.

\subsubsection{BBPO: {{Baub{\"o}ck}, {Berti}, {Psaltis},  \& 
 {{\"O}zel}} (2013) \cite{SF_2013_Baubock}}

The BBPO shape function was introduced for use in a ray-tracing code \cite{SF_2012_Baubock_Eliptical_Morsink} which makes use of the generalized quasi-Kerr metric \cite{2006CQGra..23.4167Glampedakis}, written in Boyer-Lindquist coordinates, which models spacetimes with arbitrary quadrupole moments. The shape function is computed using the Hartle-Thorne \cite{1967Hartle,1968Hartle} slow-rotation approximation, and then fit to the function
\begin{equation}
    R_{\mathrm{HT}}(\theta)\;=\;R_{0}+\xi_{2}\,P_{2}\!\bigl(\cos\theta\bigr),
\label{eq:baubock13_ht}
\end{equation}
where $R_{0}$ is the equatorial circumferential radius in the non-spinning limit and  
$\xi_{2}$ is a coefficient computed in the slow-rotation formalism \cite{SF_2013_Baubock} and depends on the dimensionless parameters $\zeta$ and $\epsilon$. As a technical point, their dimensionless parameters formally depend on the non-rotating values of mass and radius, but they argue \cite{SF_2013_Baubock} that it's possible to interchange the rotating and non-rotating mass and radius in the slow-rotation approximation.
The subscript "HT" indicates that the radius is computed within the coordinate system used in the Hartle-Thorne formalism (which differs from the Schwarzschild radial coordinate).

The coefficients for the BBPO shape function were determined by fitting to the EOSs AP4, ENG, MPA1, and MS0, which use the naming conventions established by \citet{EOS_LATTIMER_PRAKASH_2001ApJ...550..426L}. The surface defined in the Hartle-Thorne coordinate system is then  mapped \cite{SF_2013_Baubock} into the Boyer-Lindquist (BL) coordinates via  
\begin{equation}
    R_{\mathrm{BL}}(\theta)
    \;=\;R_{\mathrm{HT}}(\theta)
    -\bigl(GM/c^{2}\bigr)^{2}a^{2}\,
    \frac{\Xi(\theta)}{2R_{\mathrm{HT}}^{\,3}},
\label{eq:baubock13_bl}
\end{equation}
with  
\begin{equation}
    \begin{split}
    \Xi(\theta)=&
    \bigl(R_{\mathrm{HT}} + 2GM/c^{2}\bigr)
    \bigl(R_{\mathrm{HT}} - GM/c^{2}\bigr) \\[2pt]
    &{}- \cos^{2}\theta\,
    \bigl(R_{\mathrm{HT}} - 2GM/c^{2}\bigr)
    \bigl(R_{\mathrm{HT}} + 3GM/c^{2}\bigr)
    \end{split}
\label{eq:baubock13_xi}
\end{equation}
where $a$ is the dimensionless Kerr
spin parameter. 
An empirical fit for the spin parameter's dependence on $\zeta$ and $\epsilon$ is also found \cite{SF_2013_Baubock}. The BL radial variable reduces to the coordinate $r$ in the slow-rotation limit.

The BBPO shape function has been used  to study the effects of spot size in pulse profile modeling \cite{2015ApJ...811..144Baubock}, the effects of rotation on thermal spectra
\cite{LC_2015_Baubock_Thermal}, and in machine-learning approaches to pulse-profile modeling \cite{LC_ML_Waldrop_2025}.

The main limitation of the BBPO shape function is that it is limited to slow rotation. It is best adapted for use in codes that work in the BL coordinate system, but awkward for use in codes that are based on the Schwarzschild coordinate system.

\subsubsection{AM: AlGendy \& Morsink (2014) \cite{SF_2014_AlGendy}}

The RPPs observed by NICER have spin frequencies in the approximate range of 200 - 350 Hz, which for most masses and radii corresponds to slow rotation. This motivates the AM shape function, which attempts a higher-precision fit to the shape by restricting the spin parameter to $\epsilon \le 0.1$, which is appropriate for slow rotation and a shape function with a simpler angular dependence,
\begin{equation}
    R(\theta)\;=\;R_{\rm eq}\bigl[1+c_{2}(\epsilon,\zeta)\cos^{2}\theta\bigr],
    \label{eq:algendy_shape}
\end{equation}
where $c_{2}=c_{20}\,\epsilon + c_{21}\,\epsilon\zeta$ and $c_{20}$ and $c_{21}$ are the fit coefficients \cite{SF_2014_AlGendy}.

In the AM fits, the identity $R(\pi/2) = R_{\rm eq}$ is enforced, and neutron stars with masses $M \ge 1.0 M_{\odot}$ up to the maximum mass for the EOS were computed. The fit coefficients for AM were derived from a set of representative nuclear EOSs designed to span the plausible range of stiffness permitted by observational constraints at the time.
The EOSs include EOS BBB2  \cite{EOS_BALDO_1997A&A...328..274B}, APR  \cite{EOS_AKMAL_PhysRevC.58.1804}, ABPR1 \cite{EOS_ALFORD_2005ApJ...629..969A}, H4 \cite{EOS_LACKEY_PhysRevD.73.024021}, and the HLPS family (HLPS1, HLPS2, HLPS3) \cite{EOS_Hebeler_2013_HLPS} that represent the softest, intermediate, and stiffest EOSs consistent with then-current nuclear interaction theory.

The AM shape function is used in all of the analyses of NICER timing data of RPPs, as well as the AMXPs SAX~J1808-3658 \cite{2026MNRAS.545f1983Dorsman},  SRGA~J144459.2-604207 \cite{2026arXiv260518731Dorsman}, and burst oscillations from  XTE J1814-338 \cite{2024MNRAS.535.1507Kini} and 4U~1636-536 \cite{2025MNRAS.541...46Kini}. The AM shape function is also used in the analysis of unpulsed data of 47~Tuc~X7 \cite{2026Kazantsev} and SAX~J1810.8-2609 \cite{2020A&A...639A..33Suleimanov}. Since the AM shape function is currently used in the analysis of many neutron stars' X-ray fluxes, it is important to revisit its accuracy. In particular, since it was developed under the constraint that $\epsilon \le 0.1$, it may introduce systematic errors in the analysis of more rapidly rotating accreting neutron stars.

\subsubsection{SPYY: Silva, Pappas, Yunes, \& Yagi (2021) \cite{SF_2021_Silva}}

The shape functions introduced in the previous sections correspond to perturbations around a spherical shape using Legendre polynomials. An
alternative method \cite{SF_2021_Silva} is to perturb around the ellipsoidal function 
\begin{equation}
    R(\mu)
    = R_{\rm eq}\,
      \sqrt{\frac{1-e^{2}}{1-e^{2}\,g(\mu)}},
  \label{eq:silva21_shape}
\end{equation}
where $\mu = \cos\theta$. The function $g(\mu)$ is defined by $g(\mu)=1+a_{2}\mu^{2}+a_{4}\mu^{4}-(1+a_{2}+a_{4})\mu^{6}$ and  $e$ is the geometric eccentricity, $\,e^{2}\!=\!1-(R_{\rm pole}/R_{\rm eq})^{2}\,$. The coefficients $a_{2}$, $a_{4}$, and $e$   depend on different coefficients $c^{(y)}_{n, m}$ through the equation $y$,
\begin{multline}
y(\zeta,\epsilon)=
    c^{(y)}_{0,0}
    +c^{(y)}_{1/2,0}\,\epsilon^{1/2}
    +c^{(y)}_{1,0}\,\epsilon 
    \\[2pt]
     +c^{(y)}_{0,1}\,\zeta
    +c^{(y)}_{1,1}\,\epsilon\zeta
    +c^{(y)}_{2,0}\,\epsilon^{2}
    +c^{(y)}_{0,2}\,\zeta^{2},\qquad
    y\in\{e,a_{2},a_{4}\}.
\end{multline}
            
\citet{SF_2021_Silva} supply two separate coefficient sets, which apply for different $\epsilon$ regimes: “slow” ($\epsilon\le0.25$) and “fast” ($\epsilon\ge0.2$). To distinguish between the coefficient sets, we shall call them SPYY(s) for the slow coefficients and SPYY(f) for the fast coefficients. This split into slow and fast is straddled by the range of $\epsilon$ appropriate for stars spinning at 600 Hz, as listed in Table~\ref{tab:parameter_space_examples_v2}, which leads to some complications in its implementation.
The set of EOSs used to construct the models that went into the fit coefficients for SPYY includes FPS \citep{EOS_FPS_PhysRevLett.70.379}, SLy4 \citep{EOS_SLY4_2001A&A...380..151D}, AU \citep{EOS_AU_PhysRevC.38.1010}, UU \citep{EOS_UU_NEGELE1973298}, APR \citep{EOS_AKMAL_PhysRevC.58.1804}, and L \citep{EOS_L_1976ApJ...208..550P}.

\subsubsection{PVLS: Papigkiotis, Vardakas, Likas, \texorpdfstring{\&}{&}  Stergioulas (2025) \cite{SF_Papigkiotis_ML}}

The PVLS shape function is based on an artificial neural network (ANN) \cite{SF_Papigkiotis_ML} trained on 70 different EOSs and  $\sim4\times10^{4}$ rotating neutron star models
to predict the surface shape. Rather than being constrained to one polynomial or ellipsoidal dependence over the whole surface, the model is free to learn different behavior at different colatitudes. It has the freedom to model much more arbitrary functions across the parameter space.
The PVLS ANN-based machine learning model returns the circumferential radius $R(\mu)$ given the input parameter vector $\bm\theta^{\!*} = (|\mu|,\zeta,\epsilon,e\bigr)$, where $e$ is the eccentricity. It uses the functional form
\begin{equation}
    R(\mu)=R_{\rm pole}
        +\bigl(R_{\rm eq}-R_{\rm pole}\bigr)\,
        \hat{F}_{\bm\theta^{\!*}}\!\bigl(|\mu|,\zeta,\epsilon,e\bigr).
        \label{eq:pvls_shape}
\end{equation}
Here $\hat{F}_{\bm\theta^{\!*}}$ represents a call to the ANN model.
Since the shape function is not analytic, another ANN model is trained for the logarithmic derivative of the shape function, $d\!\log\!R/d\theta$.

In typical applications, we have the inputs $(|\mu|,\zeta,\epsilon,R_{eq})$. Notably absent is $e$.  To obtain $e$, we utilize their separate polynomial fits for the ratio $R_{\rm pole}/R_{\rm eq}$ and $e$, which are functions of our inputs. We then invoke calls to their pre-trained models. Our implementation of the PVLS shape function follows the sample Python code published with their paper \cite{SF_Papigkiotis_ML}.  
        
PVLS was trained on a comprehensive set of models derived from realistic equations of state sourced from the CompOSE database \citep{COMPOSE_Typel:2013rza, COMPOSE_Oertel:2016bki}, covering hadronic, hyperonic, and hybrid EOSs, each providing a full description of the neutron star interior from crust to core. This EOS collection encompasses seventy cold EOSs in total, which represents the largest ensemble employed for obtaining shape functions to date. The models computed include the spin parameter $\epsilon = 0 $ (non-rotating) and the non-zero range of $0.0328 \le \epsilon \le 0.9612$, as well as compactness ratios $0.0876 \le \zeta \le 0.3095$.

As is standard in machine learning, PVLS takes the set of input data (here, the computed shapes from EOSs at different spin frequencies and densities) and divides it randomly into testing and training data.  The ANN is trained on the training data and then evaluated on how well the model performs against the test data.

An advantage of the artificial neural network (ANN) approach is its ability to potentially identify and account for subtle differences not described by geometric formulas across the surface, if any exist, potentially yielding more accurate surface shapes.

In our implementation of their neural network within our modeling code, we utilized their analytical polynomial approximation for $R_{\rm pole}/R_{\rm eq}$, which is a function of $(\zeta, \epsilon)$; as in typical applications, including flux modeling, we do not exactly know the eccentricity of the neutron star a priori. However, this differs from what was implemented in their code to illustrate the accuracy of their ANN. In general, the accuracy of the PVLS shape function is limited by the accuracy of their polynomial expression for $R_{\rm pole}/R_{\rm eq}$, an issue that we will discuss in more detail in Section~\ref{sec:PVLS_polar_radius}.


\subsection{EOSs}
\label{subsec:eos_sample}

The goal of this paper is to test the performance of the 5 shape functions introduced earlier by testing whether a shape function calibrated on one set of EOSs can reproduce surfaces computed from other EOSs. The 5 shape functions were all trained on different EOS collections where all the EOSs include a hadronic component and may or may not include hyperons or a quark core.  Since we only require examples of types of errors, we don't require an extensive collection of EOSs and have chosen only 7 EOSs.  All are cold EOSs for matter in $\beta$ equilibrium, and have maximum non-rotating masses that are at least 1.97 $M_\odot$.
Four of the EOSs are taken from the CompOSE database (CompOSE labels are given below with our abbreviations in square brackets) \cite{COMPOSE_Typel:2013rza, COMPOSE_Oertel:2016bki, COMPOSE_typel2022composereferencemanual}: a hadronic EOS using a chiral SU(3) model "DS(CMF-2) with crust" [CMF2] \citep{EOS_CMF2_Dexheimer_2008}; two EOSs using the Thomas-Fermi approximation and two different density functionals, "XMLSLZ(NL3)" [NL3], and "XMLSLZ(PKDD)" [PKDD] \citep{EOS_PKDD_Xia_2022}; and a hybrid hadronic-quark EOS "KBH(QHC21A)" [QHC21A] \citep{2022ApJ...934...46K}. These 4 EOSs are moderately stiff, with $R_{1.4}$ ranging from 12.4 to 14.6 km. Of this group, only CMF2 was used in the PVLS shape function. None of these 4 EOSs were used in the other 4 shape functions. We additionally use 3 piecewise polytrope models, HLPS1-3 \citep{EOS_Hebeler_2013_HLPS}, that are consistent with nuclear experiments as of 2013 and were used in the AM shape function. The softest of these, HLPS1, is much softer than any of the EOSs used in the PVLS shape function.

We do not consider any EOSs that describe bare quark stars without hadronic crusts in this analysis. Bare quark stars are known to have shape functions that differ somewhat from the hadronic shape functions \cite{OS_SF_2007_Morsink,2026ApJ...997...55Konstantinou}. Since the 5 shape functions do not include bare quark stars, such stars are outside of the scope of this work. As a reminder, when we are testing the universality of a shape function, what we mean is quasi-universality restricted to hadronic EOSs.

\section{Shape Function Errors}
\label{sec:shape_function_errors}

One of the goals of this paper is to understand the magnitudes, morphology, and dependencies of the $R(\theta)$ errors in the various shape functions introduced in Section \ref{sec:shapes}. In this section, we present our comparisons of the shapes computed with \texttt{RNS}, $R(\theta)\equiv R_{\rm RNS}(\theta)$, and the various approximate shape functions $R_{\rm sf}(\theta)$ calculated with the same values of $M$, $R_{\rm eq}$, and $\nu$ as the \texttt{RNS} reference shape. 

Table~\ref{tab:rns_grid} summarizes the \texttt{RNS} input parameter grid used to generate the tabulated reference surfaces. The nominal grid contains $7 \times 14 \times 9 = 882$ possible \texttt{RNS} input combinations, although only combinations for which \texttt{RNS} returned a valid stellar model were retained.

\begin{table*}[!ht]
\centering
\caption{RNS input parameter grid used to generate the tabulated reference surfaces.}
\label{tab:rns_grid}
\begin{tabular}{lll}
\hline
Input axis & Values & Notes \\
\hline
EOS &
CMF2, HLPS1, HLPS2, HLPS3, NL3, PKDD, QHC21A &
 \\

Central density $\rho_c$ &
$3,4,5,6,7.5,10,12.5,15,17.5,20,22.5,25,27.5,30
$ &
$ \times 10^{14}\,\mathrm{g\,cm^{-3}}$ \\

Spin frequency $\nu$ &
$10^{-5},100,200,300,400,500,600,700,800\,\mathrm{Hz}$ &
The non-rotating limit is $10^{-5}$ Hz \\

\hline
\end{tabular}
\end{table*}

For each model, we compare the shape function surface to the corresponding \texttt{RNS} surface at the same values of $M$, $R_{\rm eq}$, and $\nu$. We define the local radius difference by
\begin{equation}
\Delta R(x)
=
R_{\rm sf}(x)-R_{\rm RNS}(x),
\qquad
x\in\{\mu,\theta\},
\label{eq:radius_difference}
\end{equation}
where $\mu=\cos\theta$. All signed errors are defined as the shape function value minus the \texttt{RNS} reference value. A positive radius error therefore means that the approximate shape function places the surface outside the \texttt{RNS} surface at that colatitude.

In the radius error plots, we show the signed percent radius error
\begin{equation}
E_R(x)
\equiv
100
\frac{
\Delta R(x)
}{
R_{\rm RNS}(x)
},
\qquad
x\in\{\mu,\theta\}.
\label{eq:percent_radius_difference}
\end{equation}
This is the quantity plotted in the colatitude-dependent radius error figures, in the $d\Omega$ surface comparison plots, and used in the universality analysis below.

\subsection{Latitudinal Morphology of the Radial Error}

\begin{figure*}[!ht]
    \centering
    \includegraphics[width=\linewidth]{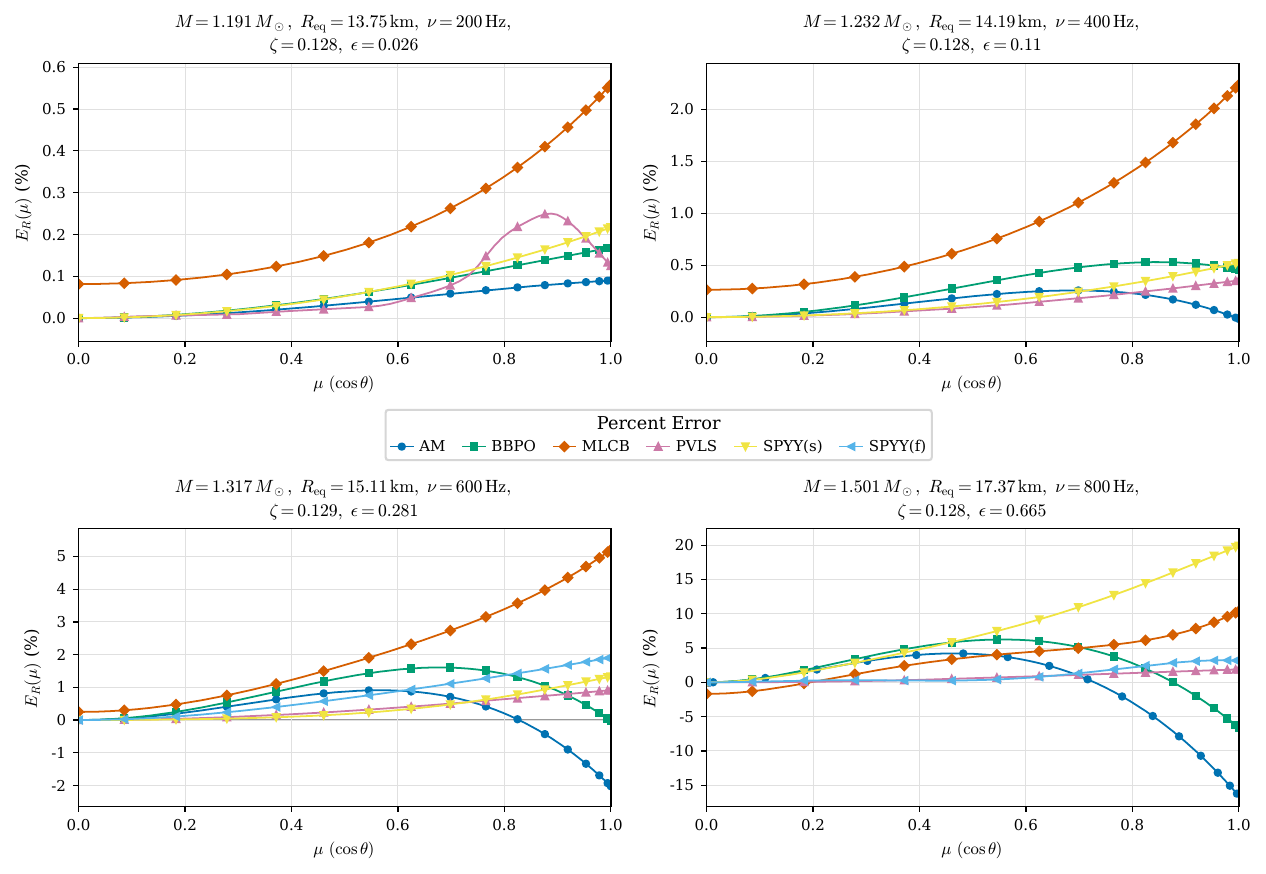}
    \caption{
    Percent error in shape functions $R(\mu)$ as a function of $\mu = \cos\theta$.
    Each panel represents one stellar model computed with \texttt{RNS} using the EOS CMF2 with a central density of $5\times10^{14}\,\mathrm{g\,cm^{-3}}$, but a different spin frequency. The equator corresponds to $\mu=0$, and the spin pole is $\mu=1$.
    Positive values for the percent error indicate that the shape function gives a larger radius than the \texttt{RNS} reference model at that colatitude. SPYY(f) is only shown where $\epsilon \geq 0.2$. This figure illustrates examples of the morphology of shape function radial errors as a function of latitude.}
    \label{fig:shape_morphology_examples}
\end{figure*}

We found three broad morphologies of shape function errors. These errors depend on the choice of shape function, but not on the chosen EOS. For illustrative purposes, we show the dependence of the shape function error as a function of $\mu$ for 4 stars constructed from the CMF2 EOS in Figure \ref{fig:shape_morphology_examples}. Each panel of Figure \ref{fig:shape_morphology_examples} shows the errors for a star with central density $5.0 \times 10^{14}\,\mathrm{g\,cm^{-3}}$ and different spin frequencies. The values of $M$, $R_{\rm eq}$, $\nu$, $\zeta$, and $\epsilon$ are shown for each panel. As shown elsewhere \cite{2022ApJ...934..139Konstantinou}, the value of $\zeta$ for stars with the same EOS and central density is approximately independent of spin. In each panel, the percent error in the function $R(\mu)$ is plotted as a function of $\mu$, where $\mu=0$ corresponds to the equator, and $\mu=1$ is the spin pole. The types of errors that are shown in Figure \ref{fig:shape_morphology_examples} are described in the next sections. These error types recur across different equations of state with different compactness and spin parameters, indicating that the colatitude dependence of the radius error depends on the structure of the shape function rather than any one EOS.

\subsubsection{Less (or More) Elliptical Than the \texttt{RNS} Reference Model}

The error in the radius of the surface and its derivative with respect to $\mu$ both grow monotonically from the equator to the spin pole, with the greatest error at the pole. The MLCB, SPYY, and PVLS (for $\epsilon > 0.04$) shape functions illustrate this type of error in all panels, with positive error at the pole. Positive error at the pole means that the approximate polar radius is too large, meaning that the shape is overall less elliptical than the reference shape. The magnitudes of the errors for the SPYY and PVLS shape functions are typically similar, with slightly smaller errors for PVLS in most cases.
 We have found a few examples of the SPYY shape function with different values of $\zeta$ that are more elliptical than the reference shape. We speculate that the SPYY shape function contains enough freedom in the $\mu$ dependence that an update could provide an approximation with much smaller errors.

\subsubsection{Bloated (or Pinched) at Mid-Colatitudes}

The error in the radius has a maximum or minimum at a mid-colatitude. The BBPO and AM shape functions for $\epsilon > 0.1$ show this behavior. Note that neither of these shape functions is meant to be used for the higher spin cases shown in the lower two panels, but we show the errors to illustrate the issue. Both of these shape functions only include contributions that are quadratic in $\mu$, which does not provide enough functional freedom to model the shape of more rapidly rotating neutron stars. However, they both provide a good description of slowly rotating neutron stars.

\subsubsection{Irregular}

The error does not have an easily described shape, possessing bumps or valleys, which do not vary in a smooth way. The PVLS shape function at small values of $\epsilon$ exhibits irregular errors, which are manifested as
the "bump" in the PVLS shape error seen in the 200 Hz
panel. While this is not a large error, it is concerning, since this is a typical spin frequency observed in RPPs. The non-zero spin training data for the PVLS shape function only included models with $\epsilon \ge 0.0328$, while this star has $\epsilon$ below the training range. This bump is seen in most models with small $\epsilon$ approximated with PVLS. This suggests that future ANNs should be trained on models with smaller values of $\epsilon$.

\subsection{Polar Radius in the PVLS Shape Function}
\label{sec:PVLS_polar_radius}

The errors in the PVLS shape function shown in Figure \ref{fig:shape_morphology_examples} are typically largest at the pole and increase in magnitude with $\epsilon$, approaching $1\%$ in some panels. This should be contrasted with the errors computed in the original paper by \citet{SF_Papigkiotis_ML}, which are generally much smaller (see, for example, their Figure 21). This discrepancy is due to how the model takes in information about the polar radius. In their PVLS analysis, \texttt{RNS} was used to compute the star's equatorial and polar radii and the eccentricity, which were then input into the ANN to predict the dependence on angle through equation (\ref{eq:pvls_shape}). Since the \texttt{RNS} polar radius is input into the equation for the radius \cite{SF_Papigkiotis_ML}, the error at the pole is typically of order $10^{-3} - 10^{-4}$ \%, and does not show much variation with spin rate.

\begin{figure}[!ht]
    \centering
    \includegraphics[width=\linewidth]{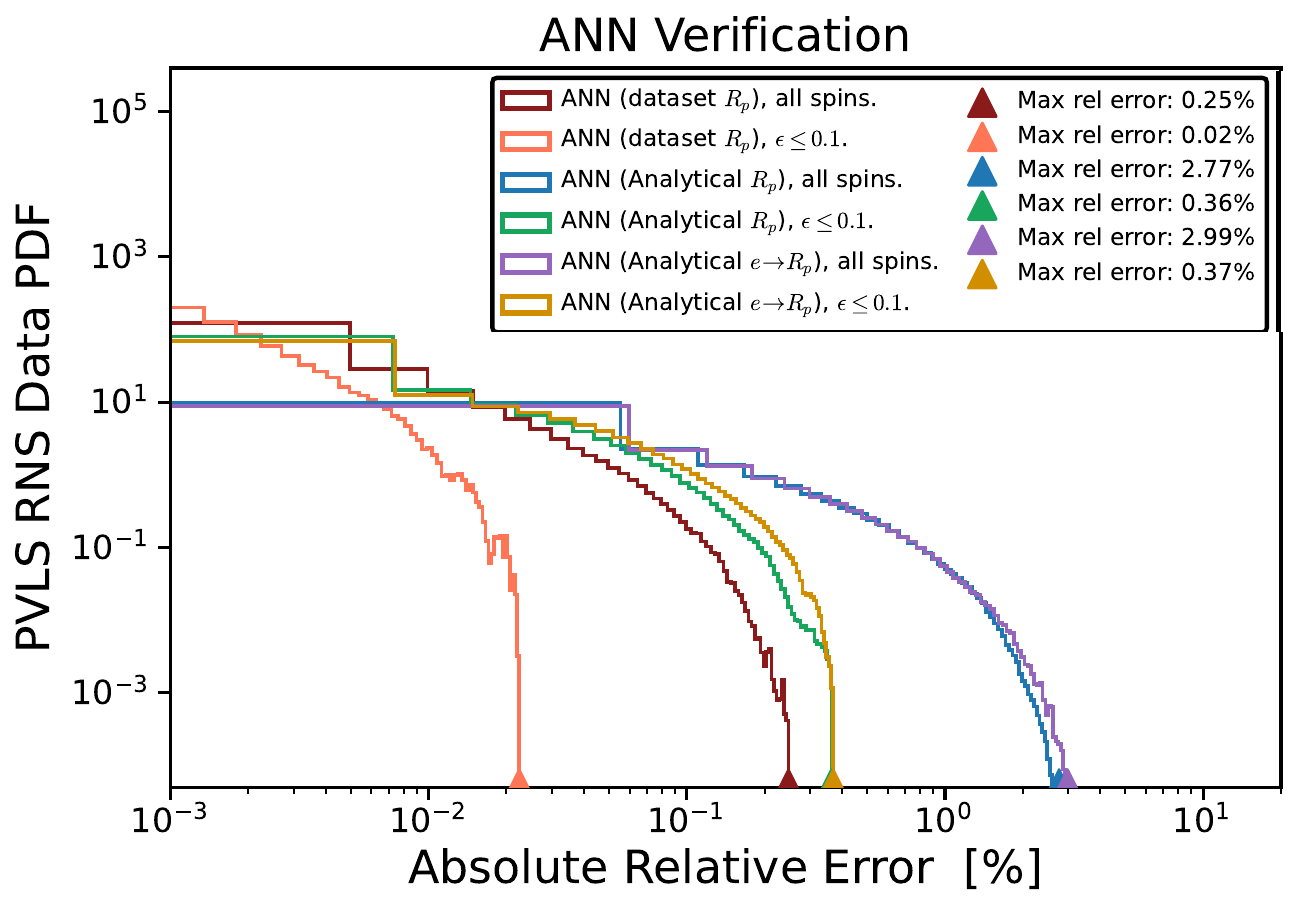}
    \caption{Histogram of the distribution of errors relative to the PVLS library \cite{SF_Papigkiotis_ML} of neutron star shapes computed using \texttt{RNS}. PDF refers to the probability density function; absolute relative error refers to the magnitude of the relative difference between the ANN's prediction and the \texttt{RNS} reference model. The binned data consist of all tabulated $\mu$ points on all surfaces of every stellar model computed by \cite{SF_Papigkiotis_ML}. Dataset $R_{\rm p}$ refers to a polar radius taken from the \texttt{RNS} reference model, analytical $R_{\rm p}$ refers to utilizing an $R_{\rm pole}$ dependent on their analytical fit for $R_{\rm p}$, and analytical $e \rightarrow R_{\rm p}$ refers to computing  $R_{\rm p}$ via their polynomial fit for eccentricity. Max rel error is the maximum relative error out of all stellar models at any point along the surface. In this paper, we are limited by the accuracy of their analytical $R_{\rm p}$.}
    \label{fig:PVLS_Error_Histogram}
\end{figure}

However, in applications such as pulse-profile modeling, the polar radius is not a known quantity. Instead, the equatorial radius and mass are free parameters, and the polar radius (or the eccentricity) has to be predicted using a universal shape function. In order to use the PVLS shape function in pulse-profile models, it is necessary to first use their polynomial fit for the polar-to-equatorial radius ratio, which depends on $\zeta$ and $\epsilon$, and has errors that grow with $\epsilon$ that can be as large as about 3\%. The error in their polynomial fit for the polar radius then dominates the error in the shape function. In
Figure \ref{fig:PVLS_Error_Histogram}, we recreate the ANN error histograms (using tan and brown colors) shown in Figure 20 of PVLS \cite{SF_Papigkiotis_ML}, using the dataset of \texttt{RNS} 
neutron star surface shapes and polar radii available in their online repository. When we make use of the PVLS shape models but use their approximate formula for $R_{\rm pole}$ instead of the \texttt{RNS} reference polar radius, the distribution of errors shifts to larger values. In particular, the maximum error over their entire dataset increases by about an order of magnitude from $ 0.25\%$ (brown triangle) to  $2.77\%$ (blue triangle). We also used their analytical fit for eccentricity to predict the polar radius and found it to be slightly less accurate than their analytical fit for $R_{\rm pole}$, as shown by the purple histogram and triangle. The tan, green, and gold histograms show the distribution of errors when the restriction $\epsilon < 0.1$ is applied, where tan shows the original PVLS histogram, and the green and gold histograms show the increase in errors when the PVLS analytical approximations for $R_{\rm pole}$ and $e$ are used, respectively. The maximum error for the low-spin dataset also increases by an order of magnitude from $0.02\%$ (tan triangle) to  $0.36\%$ (green triangle). For reference, with the same dataset, SPYY(s) has a maximum relative error of $1.12\%$ at $\epsilon \leq 0.1$ and a maximum relative error of $5.12\%$ across the whole dataset; AM has a maximum relative error of $0.56\%$ at $\epsilon \leq 0.1$ and a maximum relative error of $44.11\%$ across the whole dataset. 

The jump in the magnitude of the errors in the PVLS shape function shown in Figure~\ref{fig:PVLS_Error_Histogram} illustrates that the key factor limiting the accuracy is the model for the polar radius given the equatorial radius, spin parameter, and compactness. If we were able to obtain a perfectly accurate model for the polar radius, then our remaining errors in the mid-latitudes would only be on the order of $0.25\%$.

\subsection{Errors in the Derivative of the Shape Function}
\label{sec:E_r'_errors}

The derivative of the shape function with respect to colatitude determines the direction of the local surface normal. In the OS approximation, it affects the surface-slope factor $q$ and the angle $\tau$ between the radial direction and the surface normal defined in equations (\ref{eq:tau}) and (\ref{eq:surface_slope_q}). This error then propagates into the differential solid angle and flux.

\begin{figure*}[!ht]
    \centering
    \includegraphics[width=\linewidth]{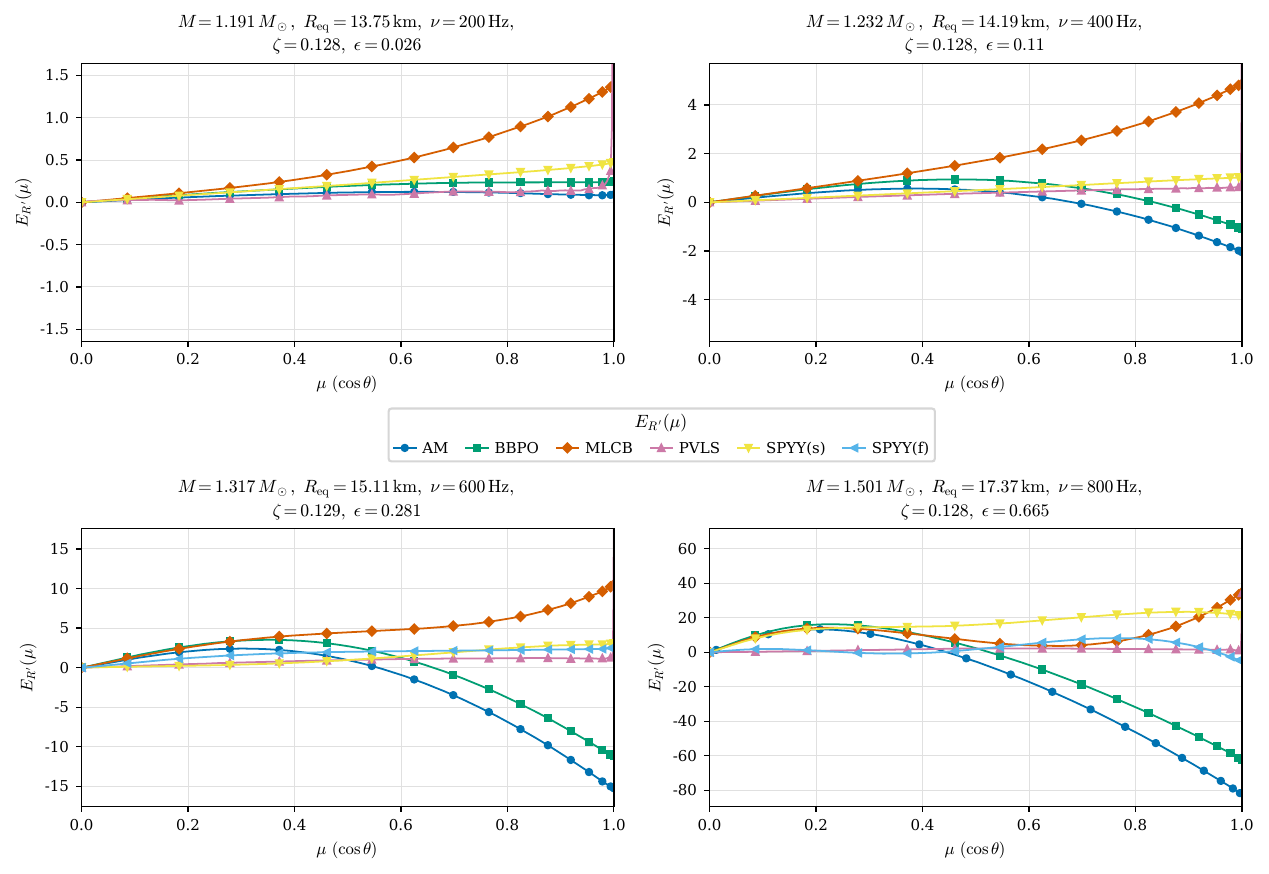}
    \caption{
    Signed derivative error $E_{R'}(\mu)$ for the shape functions as a function of $\mu=\cos\theta$. Each panel represents one stellar model computed with \texttt{RNS} using the EOS CMF2 with a central density of $5\times10^{14}\,\mathrm{g\,cm^{-3}}$, but a different spin frequency. The equator corresponds to $\mu=0$, and the spin pole is $\mu=1$. Positive values indicate that the shape function has a larger slope than the \texttt{RNS} surface at that colatitude.}
    \label{fig:dr_dmu_errors_morphology_examples}
\end{figure*}

For illustrative purposes, Figure~\ref{fig:dr_dmu_errors_morphology_examples} shows the derivative error as a function of $\mu$ for the same four example neutron stars shown in Figure~\ref{fig:shape_morphology_examples}. Throughout the remainder of this paper, a prime denotes differentiation with respect to the displayed argument, so $R'(\mu)\equiv dR(\mu)/d\mu$ and $R'(\theta)\equiv dR(\theta)/d\theta$.

We define the signed derivative error by
\begin{equation}
E_{R'}(x)
\equiv
100
\frac{\Delta R'(x)}{R_{\rm RNS}(x)}
=
100
\frac{
R'_{\rm sf}(x)-R'_{\rm RNS}(x)
}{
R_{\rm RNS}(x)
},
\;
x\in\{\mu,\theta\}.
\label{eq:derivative_error_definition}
\end{equation}
Figure~\ref{fig:dr_dmu_errors_morphology_examples} uses $x=\mu$, while the later surface maps use $x=\theta$. We normalize by the radius rather than its derivative to avoid uninformative division-by-zero divergences. Since $E_R$ is itself normalized by $R_{\rm RNS}(\mu)$, $E_{R'}$ is not the derivative of $E_R$. Positive values of $E_{R'}$ indicate that the shape function has a larger slope than the \texttt{RNS} surface.

The derivative errors broadly follow the same classes seen in the radius errors. Smooth one-sided radius errors have consistently positive or negative derivative errors. Mid-colatitude bumps or valleys in $R(\mu)$ appear as sign changes in the derivative error. Irregular radius errors, such as those seen with PVLS at low $\epsilon$, produce correspondingly irregular derivative errors.

The AM, BBPO, MLCB, and SPYY shape functions are analytic functions of $\mu$, so the derivative $R'(\mu)$ is simple to compute. 
\new{PVLS provides a secondary neural network to predict the radius derivative  \cite{SF_Papigkiotis_ML}. Its result need not match the numerical derivative of the radius returned by the radius neural network. In our comparison of the two PVLS radius derivatives, we found that the differences between them are typically less than $1\%$, except in regions where one neural network produces errors specific to it. For example, there is the bump error for the radius ANN (where the difference can reach $5\%$) or the spike error for the derivative ANN that will be discussed in the next paragraph.}
We use the result of the PVLS secondary neural network to compute $R'(\mu)$. 

A careful inspection of the top-left panel of Figure \ref{fig:dr_dmu_errors_morphology_examples} reveals that the PVLS error shoots upwards as $\mu$ approaches $1.0$. The same effect may be observed in subsequent $d\Omega$ plots (shown in later figures) which display $E_{R'}(\theta)$. In the original PVLS paper \cite{SF_Papigkiotis_ML}, they utilized 521 samples uniformly spaced in $\mu$, which leaves a gap between $\mu = 519/520$ and $\mu = 1.0$, corresponding to $ 0^\circ \leq \theta \lesssim 3.55^\circ$. This gap is not probed by their training data, ANN, or testing, and subsequently no issues are shown in their paper. However, we found that within this gap, their derivative ANN consistently produces a sharp spike in $E_{R'}$ on the order of $2\%$. Our subsequent results may show the full spike or only part of the spike depending on our sampling of the region near the pole.

The types of errors in the derivatives shown in Figure~\ref{fig:dr_dmu_errors_morphology_examples} recur across a broad range of EOSs, compactness, and spin parameters. From our larger collection of models, we note that the types of errors in radius and radius derivative depend on the choice of shape function, not on the EOS, which lends credence to the concept of universality.

\subsection{Variation of the Shape Error Across the Parameter Space}
\label{sec:shape_variation}

We now investigate the dependence of the errors in the five shape functions on the dimensionless compactness and spin parameters $\zeta$ and $\epsilon$ using the rotating neutron star models generated with \texttt{RNS} shown in Table~\ref{tab:rns_grid}. To simplify the analysis, for each model computed and shape function tested, we introduce a signed extremal radius error, ${\rm MaxE}_{R}$. Let $\mu_{\rm max}$ be the value of $\mu=\cos\theta$ at which the absolute percent radius error is largest,
\begin{equation}
\mu_{\rm max}
=
\operatorname*{arg\,max}_{0\leq\mu\leq1}
\left|
E_R(\mu)
\right| .
\end{equation}
We define the signed extremal radius error 
\begin{equation}
{\rm MaxE}_{R}
=
E_R(\mu_{\rm max}) .
\label{eq:max_radius_error}
\end{equation}
The extremal radius error is the signed percent radius difference at the colatitude where the radius error lies furthest from zero. A positive value means that the shape function is too large at the worst-error colatitude.

\begin{figure*}[!ht]
    \centering
    \includegraphics[width=0.45\linewidth]{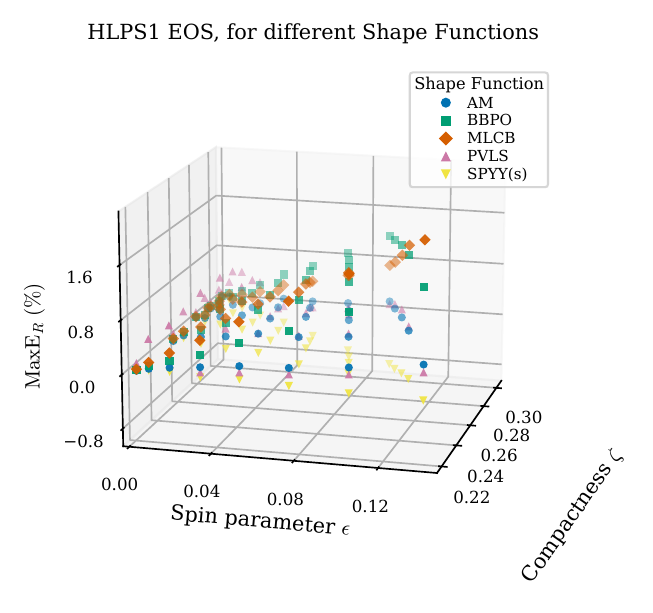}
    \includegraphics[width=0.45\linewidth]{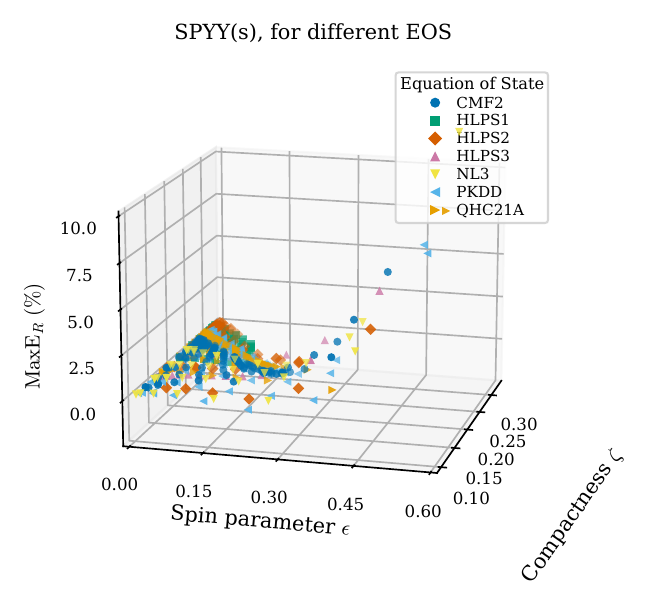}
    \caption{
    Representative examples of the dependence of the extremal radius error on compactness and spin parameter. The left panel shows a zoomed-in view of the errors relative to the HLPS1 EOS, illustrating how different shape functions possess different error morphologies over the same region of parameter space. The right panel shows the errors of SPYY(s) \cite{SF_2021_Silva} relative to our set of EOSs, illustrating how the shape function's errors appear to fall on the same smooth surface regardless of EOS. Positive values indicate that the shape function overestimates, while negative values indicate an underestimate. In both panels, the transparency of a point corresponds to the compactness of the neutron star.}
    \label{fig:Shape_Error_Example}
\end{figure*}

The left panel of Figure~\ref{fig:Shape_Error_Example} shows the variation in the extremal radius error as a function of compactness and spin parameter (for small $\epsilon$) for the five shape functions for one particular choice of EOS (HLPS1). The shape functions have different error patterns and dependence on the spin parameter and compactness. This illustrates that each shape function behaves differently within the parameter space. For brevity, we only show models computed with one EOS, but these patterns apply to all of the EOSs that we tested. 

The right panel of Figure~\ref{fig:Shape_Error_Example} illustrates the extremal radius error for one specific shape function, SPYY(s), for all 7 EOSs. In this plot, the extremal radius errors appear to form a surface in the $\epsilon-\zeta$ parameter space, independent of the choice of EOS. This surface, formed by the errors, hints at the validity of the assumption underlying universality. 
If the neutron star shape were not universal, the errors for different EOSs would show random scatter for models with the same value of $\zeta$ and $\epsilon$. Since the errors for any shape function show structure, we hypothesize that a better empirical fit is possible. 
The same extremal radius error is shown for three of the shape functions in panel (a) of Figure~\ref{fig:universality_surface}.
We explore the possibility of better empirical fits in the next section.

Figure \ref{fig:PVLS_Error_Histogram} may appear to suggest that large relative errors are unlikely. This is due to the different error summaries being emphasized. Figure~\ref{fig:PVLS_Error_Histogram} presents the distribution of pointwise errors over all sampled colatitudes ($521$ points over $\mu \in [0, 1]$) and all stellar models, in which a small number of local high-error points contribute only weakly to the overall distribution.  The maximum error defined in this section emphasizes the largest difference on the surface of each particular star.

\subsection{Universality Analysis}

In Section~\ref{sec:shape_variation}, we found that the extremal radius error for any shape function appears to depend on the spin and compactness parameters. This structure suggests that the shape has an approximately universal dependence on $\epsilon$ and $\zeta$ and that it could be possible to construct a corrected shape function with smaller errors. 
We can quantify this idea by fitting a simple correction surface $R_{\rm c}(\mu,\zeta,\epsilon)$ to the error $\Delta R(\mu)$ at each value of $\mu$. We assume a simple polynomial correction to the radius error with the form 
\begin{equation}
R_{\rm c}(\mu,\zeta,\epsilon)
=
\mu
\sum_{a\in\{1,3\}}
\sum_{b+c\leq 3}
C_{abc}\,
\mu^{a}\zeta^{b}\epsilon^{c},
\label{eq:R_c}
\end{equation}
where the restriction $a\in\{1,3\}$ preserves reflection symmetry through the equatorial plane. The prefactor of $\mu$ ensures that the correction surface vanishes at the equator, so we assume that $R_{\rm sf}$ has been matched to the correct equatorial radius, as is true for all of the shape functions except MLCB. The coefficients $C_{abc}$ are found by linear regression.

Given the best-fit correction surface, the corrected shape function is
\begin{equation}
R_{\rm sf,c}(\mu)
=
R_{\rm sf}(\mu)-R_{\rm c}(\mu,\zeta,\epsilon).
\label{eq:corrected_shape_function}
\end{equation}
The remaining radius error is
\begin{equation}
\Delta R_{\rm c}(\mu)
=
R_{\rm sf,c}(\mu)-R_{\rm RNS}(\mu)
=
\Delta R(\mu)-R_{\rm c}(\mu,\zeta,\epsilon).
\label{eq:corrected_radius_error}
\end{equation}
The corrected percent radius error is
\begin{equation}
E_{R,\rm c}(\mu)
=
100
\frac{
\Delta R(\mu)-R_{\rm c}(\mu,\zeta,\epsilon)
}{
R_{\rm RNS}(\mu)
}.
\label{eq:corrected_percent_radius_error}
\end{equation}
Extremal errors in the corrected surface are defined analogously to Eq.~\eqref{eq:max_radius_error}.

We focus on fitting the errors for 3 of the shape functions: AM, SPYY(s), and PVLS. The AM shape function was examined due to its use in NICER analyses; SPYY(s) since it has a simple geometric form; and PVLS since it was created using machine learning with the largest neutron star surface dataset to date. Furthermore, PVLS typically has the lowest scale of overall error. We don't provide the values of the fit coefficients, since this is only a "proof-of-principle" done with a very limited number of EOSs.
Our fits to the errors should be thought of as an upper bound on the errors produced by a hypothetical better shape function. Most shape functions utilize a more sophisticated description of the surface $R(\theta)$, so it may be possible to construct a shape function with smaller errors than what we show here. 

\begin{figure*}[!hbt]
    \centering
    \includegraphics[width=\linewidth]{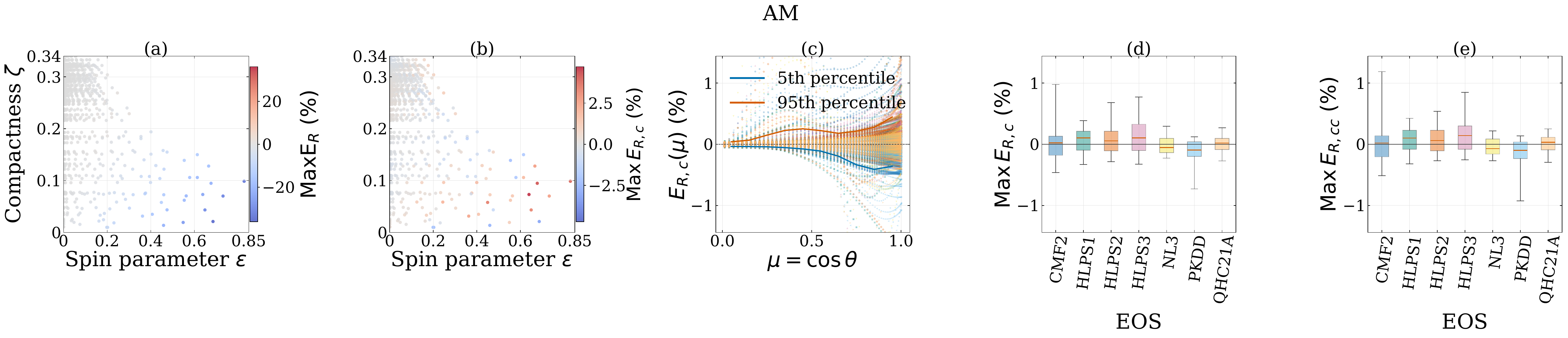}
    \includegraphics[width=\linewidth]{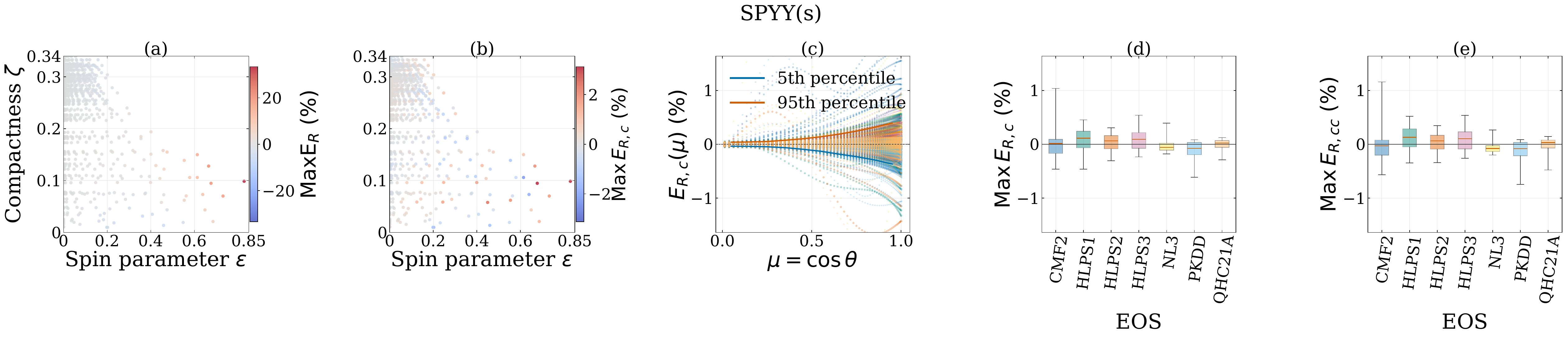}
    \includegraphics[width=\linewidth]{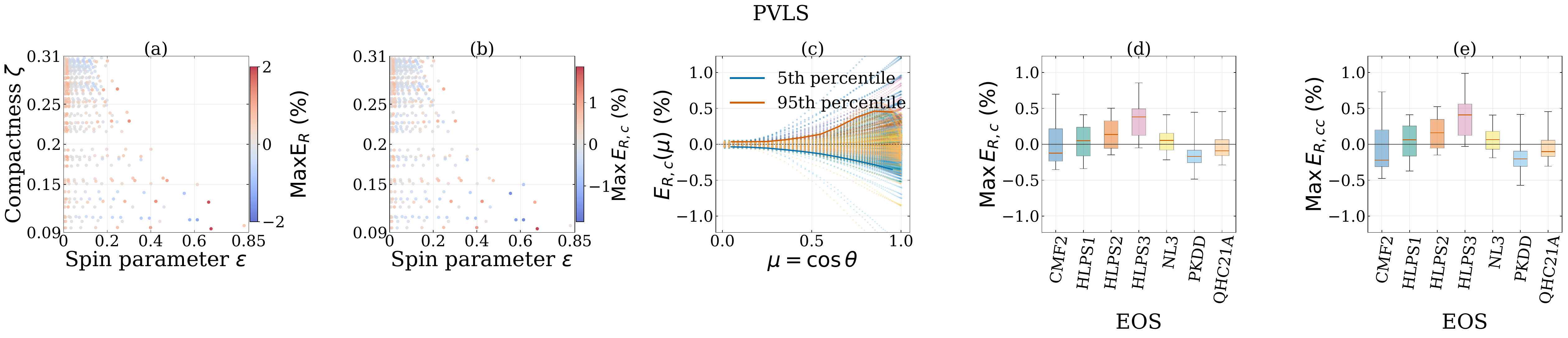}
    \caption{
    Empirical all-EOS correction surface fits to the radius errors of AM, SPYY(s), and PVLS. For each shape function, panel (a) shows the extremal radius error as a function of compactness and spin. Panel (b) shows the corrected extremal error after fitting $R_{\rm c}(\mu,\zeta,\epsilon)$ to the full EOS sample. Panel (c) shows the corrected error as a function of colatitude, with the 5th percentile marked in blue and the 95th percentile marked in orange. The points in panel (c) have the same colors as the EOSs identified in panel (d). Panel (d) shows the distribution of corrected errors by EOS for the all-EOS fit using $R_{\rm c}$, while panel (e) shows the corresponding cross-verification distribution obtained using $R_{\rm cc}$, fitted without the EOS being tested. In the box plots in panels (d) and (e), the boxes span the 25th to 75th percentiles, the central line marks the median, and the whiskers extend to the 5th and 95th percentiles. If most of the error can be modeled by a single smooth function of $(\mu,\zeta,\epsilon)$, then we may obtain an upper bound on the remaining radius error unexplained by the correction surface.}
    \label{fig:universality_surface}
\end{figure*}

Our results are presented in Figure \ref{fig:universality_surface}, with each horizontal row representing the errors in each shape function. Panel~(a) illustrates the extremal radial error dependence on $\epsilon$ and $\zeta$ for the given shape function. Note that the errors in the AM and SPYY(s) shape functions are very large because these shape functions were derived only using stars with slow rotation ($\epsilon \le 0.1$ for AM and $\epsilon \le 0.25$ for SPYY(s)). The training data for PVLS and SPYY shape functions are restricted to $\zeta \ge 0.08$. The AM shape function restricted its models to $M \ge 1.0 M_\odot$, which leads to a similar restriction on $\zeta$. In the case of PVLS, the errors for very small (and most likely nonphysical) $\zeta$ are larger, as would be expected since small $\zeta$ values were not part of their training data. Similarly, larger errors for $\epsilon \le 0.03$ can be seen as a red vertical strip in the PVLS panel, which is also due to sampling models outside the range of parameters used in the training data set. 

Panel (b) of Figure~\ref{fig:universality_surface} illustrates the corrected extremal radius error found after fitting $R_{\rm c}(\mu,\zeta,\epsilon)$ to $\Delta R$.
In the case of AM and SPYY(s), we obtain an order of magnitude decrease in the scale of their errors. In the case of PVLS, the scale of the error barely decreases. For large $\epsilon$, the remaining error in the corrected shape functions for AM and SPYY(s) shown in panel (b) has an approximately random distribution of positive and negative error residuals that don't appear to have any structure in the $\zeta$ dimension. The increase in error with $\epsilon$ suggests that a polynomial with higher powers in $\mu$ and $\epsilon$ is probably required for these cases. The residual errors for PVLS still show structure in the $\epsilon$ dimension, suggesting that the polynomial approach taken here may not be sufficient to correct for the ANN. Alternatively, we may speculate that we are approaching a floor in the assumption of universality, that shape functions might not be able to model neutron star surface shapes any more accurately. Further research is required to ascertain this. In the case of PVLS, it would probably be more useful to create a new training data set that provides more very slowly rotating models, to correct for the errors seen near $\epsilon \le 0.03$.

Panel (c) illustrates the dependence of the corrected error $E_{R,\rm c}(\mu)$ on $\mu$, with colors representing the EOSs as defined in panels (d) and (e). The majority of the remaining error continues to be towards the poles, but the bulk of the error (5th to 95th percentile) lies within $\pm 0.5\%$ for all three shape functions. Only a tiny percentage of the cases have errors approaching the ends of the $\pm 4\%$ range (which correspond to the most rapidly rotating models). Notably, all three shape functions are comparable in their 5th and 95th percentile ranges.

The boxplots in panel (d) illustrate the extremal corrected errors for individual EOSs. The red line in each box is the median, the edges of the box are the 25th and 75th percentiles, and the whiskers stretch out to the 5th and 95th percentiles. For most cases, the 25th to 75th percentile ranges are smallest and are closest to zero for SPYY. The median values show the largest variation for PVLS, where for a couple of EOSs, the 25th to 75th percentile range does not include zero. The contrast is most obvious between HLPS3 and PKDD, the former of which skews towards positive errors and the latter towards negative errors for its 25th to 75th percentile range. This contrast in their skew is seen to a lesser extent with the other shape functions. It is unclear whether this hints at a limit to the quasi-universality assumption, or if this feature emerges specifically due to the shape function. Furthermore, its 25th to 75th percentile spread is of comparable or larger scale to that of the other shape functions, despite its overall largest outliers being smaller than the other two shape functions.


The investigation so far utilized all EOSs to fit $R_{\rm c}$. However, the assumption of universality suggests that a shape function fitted for one set of nuclear equations of state will be generalizable across other EOSs. For this cross-verification fit, the corrected percent radius error is
\begin{equation}
E_{R,\rm cc}(\mu)
=
100
\frac{
\Delta R(\mu)-R_{\rm cc}(\mu,\zeta,\epsilon)
}{
R_{\rm RNS}(\mu)
}.
\label{eq:cross_corrected_percent_radius_error}
\end{equation}
In panel (e), we fit $R_{\rm cc}$ to every other equation of state except for the chosen equation of state, and then examine the extremal error for the given equation of state in the same way. We observe minimal differences between these results and those of (d), supporting the hypothesis of universality in the sense of generalization of the shape function across different EOSs, but perhaps up to a limit, as seen by the skew of the errors relative to zero as those offsets persist.

These results suggest that it would be possible to construct a more accurate version of SPYY using a fit with higher-order polynomials and a much larger collection of EOSs than were used in the original paper. 

\section{Impact of Shape Function Errors on Differential Solid Angle}
\label{sec:solid_angle_errors}

In Section \ref{sec:shape_function_errors}, we computed the errors in the surface shape $R(\theta)$ and its derivative. However, neither quantity is directly observable. We observe the flux from a patch of the star, which is proportional to the differential solid angle, $d\Omega$, subtended by the patch. In the OS approximation, errors in the shape and its derivative propagate into the differential solid angle and the flux through Eq.~\eqref{eq:domega_split}. Here, we examine the $d\Omega$. \new{Note that the errors in this section are calculated for infinitesimal surface elements. For an extended emitting area, the flux must be integrated over the visible surface area. This weights the flux and the error by the local $d\Omega$, which suppresses the contribution of errors from the limb.}

For each visible surface element, we define the local differential solid angle difference by
\begin{equation}
\Delta(d\Omega)
=
d\Omega_{\rm sf}-d\Omega_{\rm RNS},
\label{eq:domega_difference}
\end{equation}
where $d\Omega_{\rm sf}$ is the solid angle subtended by the spot on the star using one of the approximate shape functions and $d\Omega_{\rm RNS}$ is the solid angle computed using the \texttt{RNS} reference shape at the same location on the star. 
The signed percent difference is defined by 
\begin{equation}
E_{d\Omega}
=
100
\frac{
\Delta(d\Omega)
}{
d\Omega_{\rm RNS}
} .
\label{eq:percent_domega_difference}
\end{equation}

\subsection{Analytic Approximation to the Differential Solid Angle Error}
\label{subsec:analytic_differential_solid_angle}

We now derive a simple analytical approximation to the differential solid angle error in order to aid the understanding of how errors in the shape function affect the error in the differential solid angle. In the derivation of this approximation, we treat the following quantities as exact: the angular coordinates $\theta$ and $\phi$ of the hot spot on the surface of the star, the observer's inclination angle $i$, and the distance to the star $D$. This allows us to focus only on the errors introduced by the shape function. Throughout this subsection, unlabeled quantities such as $R(\theta)$, $R'(\theta)$, and $d\Omega$ refer to the quantities defined by the OS geometry, rather than to any particular numerically computed stellar model.

Based on the spherical trigonometric definitions used in the OS approximation given in Eqs.~\eqref{eq:psi} and \eqref{eq:lambda}, the angles $\psi$ and $\lambda$ are known exactly if $\theta$, $\phi$, and $i$ are known exactly.

Although the bending angle $\psi$ is known exactly, the angle $\alpha$ between the photon's initial direction and the radial direction depends implicitly on the radial location of the emitting region through Eqs.~\eqref{eq:psi_integral} and \eqref{eq:alpha}. We can gain a more intuitive understanding of this dependence by employing the Beloborodov approximation \cite{dCosAlpha_dCosPsi_Approx_Relation}, which relates $\psi$ and $\alpha$ through the expansion
\begin{equation}
    1 - \cos \psi = \frac{1}{1-\frac{2M}{R}} (1 - \cos \alpha)  + O(1-\cos \alpha)^3. 
    \label{eq:belob}
\end{equation}
Since the coefficient of the term of order $(1-\cos\alpha)^3$ is very small \cite{dCosAlpha_dCosPsi_Approx_Relation}, this approximation is surprisingly good even for photons emitted in directions far from the radial direction. Its accuracy as a function of compactness and emission angle is shown in Figure~2 of Ref.~\cite{dCosAlpha_dCosPsi_Approx_Relation}. We found that the error in the bending angle $\psi$ falls below $5\%$ for local compactness $M/R(\theta)\leq 0.2317$, and 
the error in the inferred emission angle $\alpha$ falls below $5\%$ for $M/R(\theta)\leq 0.25$. Although we do not use the Beloborodov approximation in our numerical calculations, we use it here to derive an analytical approximation to the solid angle error.

Differentiating Eq.~\eqref{eq:belob}, we find
\begin{equation}
\frac{1}{1-\frac{2M}{R}}
\left|
\frac{\partial\cos\alpha}{\partial\cos\psi}
\right|_R
= 1 + O\!(1-\cos\alpha)^2,
\label{eq:beloborodov_jacobian}
\end{equation}
so, within this level of approximation, the product of the self-lensing and Jacobian factors appearing in Eq.~\eqref{eq:domega_split} is constant and does not contribute to the error. 
The fractional error in the differential solid angle in this approximation is then
\begin{equation}
    \frac{\Delta(d\Omega)}{d\Omega}
    \simeq 2\frac{\Delta R(\theta)}{R(\theta)} + \frac{\Delta\mathcal{P}}{\mathcal{P}},
    \label{eq:domega_error_combined_projection}
\end{equation}
where the projection term ${\mathcal{P}}$
is the combination of terms
\begin{equation}
    \mathcal{P} \equiv \frac{\cos\sigma}{\cos\tau}
    = \cos\alpha + \tan\tau \sin\alpha \cos\lambda .
    \label{eq:combined_projection_factor}
\end{equation}
The first term in Eq.~\eqref{eq:domega_error_combined_projection} gives the simple intuitive result that an error in the radial distance leads to an error in the surface area at that point.        Errors in the angles $\alpha$ and $\tau$ both contribute to the error in the projection term.

Within the Beloborodov approximation, the error in the angle $\alpha$ follows from differentiating Eq.~\eqref{eq:belob} with respect to $R$,
\begin{equation}
    \Delta\cos \alpha
    \simeq
- \frac{2M}{R}
    \frac{1}{1-\frac{2M}{R}}
    (1 - \cos \alpha)
    \frac{\Delta R}{R},
    \label{eq:deltaalpha_app_explicit}
\end{equation}
where $R$ and $\Delta R$ are evaluated at the latitude of the spot.

The error in the angle $\tau$ does not require any approximation and follows directly from Eqs.~\eqref{eq:tau} and \eqref{eq:surface_slope_q},
\begin{equation}
    \Delta \tau
    =
    \cos^2\tau
    \left[
        -
        \frac{1-\frac{M}{R}}
        {1-\frac{2M}{R}}
        \tan\tau
        \frac{\Delta R}{R}
        +
        \frac{1}
        {\sqrt{1-\frac{2M}{R}}}
        \frac{\Delta R'(\theta)}{R}
    \right].
    \label{eq:delta_tau_exact}
\end{equation}
However, the angle $\tau$ is always small, so the error in $\tau$ is well approximated by
\begin{equation}
    \Delta \tau
    \simeq
    \frac{1}
    {\sqrt{1-\frac{2M}{R}}}
    \frac{\Delta R'(\theta)}{R}
    +
    O\left(
        \tau\frac{\Delta R}{R}
    \right)
    +
    O\left(
        \tau^2\frac{\Delta R'(\theta)}{R}
    \right),
    \label{eq:delta_tau_small_tau}
\end{equation}
so that the error in the angle $\tau$ between the radial and normal directions is dominated by the error in the derivative of the shape function.

The angular dependence of these errors differs. Since the surface is axisymmetric, $R$, $R'$, $\Delta R$, and $\Delta R'$ depend only on $\theta$. The error $\Delta\tau$ therefore also depends only on $\theta$. In contrast, $\Delta\cos\alpha$ depends on both $\theta$ and $\phi$ through $\alpha$, since $\alpha$ is determined by the bending angle $\psi(\theta,\phi,i)$ and the local radius $R(\theta)$. The projection factor $\mathcal{P}$ combines $\alpha$, $\lambda$, and $\tau$, so its error generally depends on both $\theta$ and $\phi$, even though the underlying shape function errors depend only on $\theta$.

The final result for the error in the projection term is
\begin{eqnarray}
    \frac{\Delta\mathcal{P}}{\mathcal{P}}
    &\simeq&
    \frac{1}{\cos\sigma}
    \left[ 
        -
        \frac{2M}{R}
        \frac{1}{1-\frac{2M}{R}}
        (1-\cos\alpha)
        \frac{\Delta R}{R} \right.
        \\
        & & \left. +
        \cos\lambda\sin\alpha
        \frac{1}{\sqrt{1-\frac{2M}{R}}}
        \frac{\Delta R'(\theta)}{R}
    \right]
    +O(\tau).\nonumber
    \label{eq:projection_error_analytic}
\end{eqnarray}
Since $M/R$ is typically small, the error in the projection term is dominated by the error in the derivative of the radius. The sign of this error depends on the sign of $\cos\lambda$, since all of the other terms are positive.

The final analytical approximation for the percent error in the differential solid angle is
\begin{eqnarray}
    E_{d\Omega}
    &\simeq&
    2E_R
    +
    \left[
        -
        \frac{2M}{R}
        \frac{1}{1-\frac{2M}{R}}
        (1-\cos\alpha)
        \frac{1}{\cos\sigma}
    \right]E_R \nonumber \\
   & + &
    \left[
        \cos\lambda
        \frac{1}{\sqrt{1-\frac{2M}{R}}}
        \frac{\sin\alpha}{\cos\sigma}
    \right]E_{R'}.
    \label{eq:domega_percent_error_analytic}
\end{eqnarray}
The projection error term is small for photons emitted near the radial direction ($\alpha \rightarrow 0$), where the error in the differential solid angle will be close to twice the magnitude of the error in the shape function.
At the limb, where $\cos\sigma \rightarrow 0$, the error from the projection term diverges since the solid angle approaches zero at the limb.  
For photons emitted in other directions, there are contributions from both the errors in the shape function and its derivative that depend on the angular location of the emitting region on the star (through the dependencies of $\alpha$, $\sigma$, and $\lambda$ on $\theta$ and $\phi$)
and on the compactness of the star. 

The sign of $\cos\lambda$ makes an important contribution to the magnitude and sign of the overall error in the solid angle. Since the dependence of $\cos\lambda$ on the values of $\phi$, $\theta$, and $i$ is nontrivial, we show its values for some special cases in Table \ref{tab:coslambda}.

\begin{table}[!ht]
    \centering
    \begin{tabular}{|c|c|c|c|c|}
    \hline
         $i$ & $\phi$ & $\theta$ & $\psi$ & $\cos \lambda$ \\
         \hline
         0 & $\; 0 \le \phi \le 2 \pi \;$ & $\;0 \le \theta \le \pi\;$ & $\theta$ & +1 \\
         \hline
         $\pi/2$ & 0 & $\theta < \pi/2 $ & $\pi/2 - \theta$ & $-1$ \\
        $\pi/2$ & 0 & $\theta > \pi/2 $ & $\theta - \pi/2$ & $+1$ \\
        $\pi/2$ & $\pi$ & $\theta < \pi/2 $ & $\theta + \pi/2$ & $+1$ \\
        $\pi/2$ & $\pi$ & $\theta > \pi/2 $ & $3\pi/2 - \theta$ & $-1$ \\
        \hline
        $0 < i < \pi/2$ &   0 &  $\theta < i $ & $i-\theta$ & $-1$ \\
        $0 < i < \pi/2$ &   0 &  $\theta > i $ & $\theta-i$ & $+1$ \\
        $0 < i < \pi/2$ &  $\pi$ & $\theta < \pi - i $ & $\theta+i$ & $+1$ \\
        $0 < i < \pi/2$ &  $\pi$ & $\theta > \pi - i $ & $2\pi -(\theta+i)$ & $-1$ \\
        \hline
    \end{tabular}
    \caption{Values of the bending angle $\psi$ and the cosine of the interior angle $\lambda$ for the full range of inclination and spot colatitude $\theta$, as well as for special values of spot longitude $\phi$. The longitude of the observer is $\phi = 0$, while $\phi=\pi$ corresponds to longitude on the side of the star opposite to the observer.}
    \label{tab:coslambda}
\end{table}


\subsection{Examples of Differential Solid Angle Error Patterns}

In this section, we provide some examples to illustrate the error in the differential solid angle, and hence its contribution to the flux. The solid angle error depends on
 the errors in the radius and the radius derivative, the observer's inclination, as well as the stellar parameters $\zeta$ and $\epsilon$. Since the dependencies depend on the position of the emitting area on the star, and are fairly complex, we focus on a few examples that best illuminate the types of problems that occur. 

\begin{figure*}[!ht]
    \centering
    \includegraphics[width=1.0\linewidth]{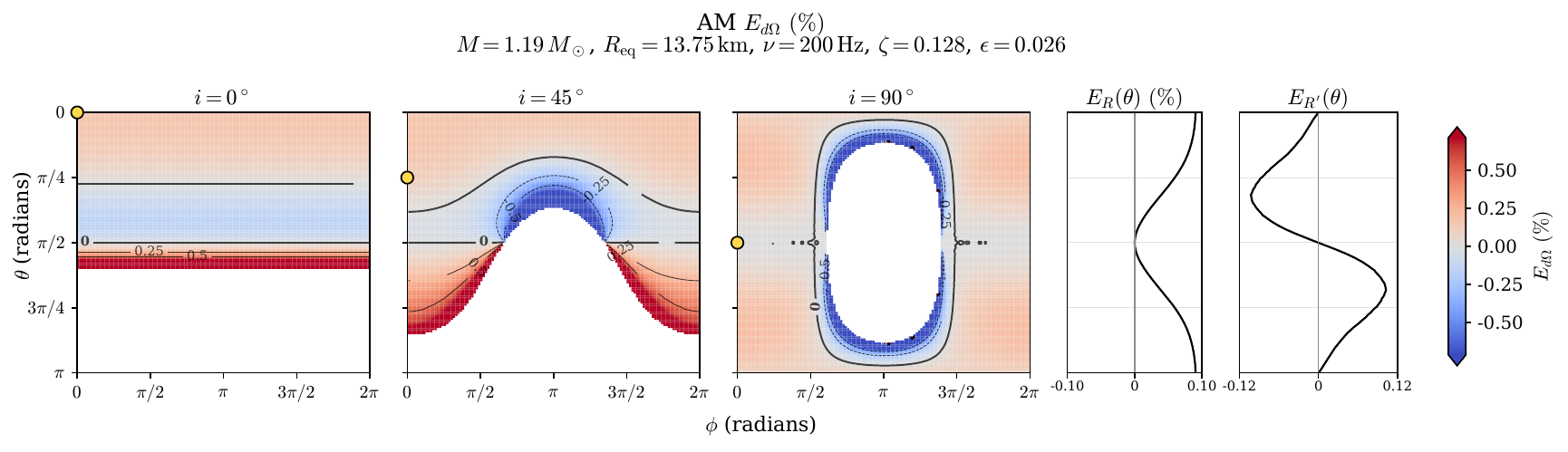}
    \includegraphics[width=1.0\linewidth]{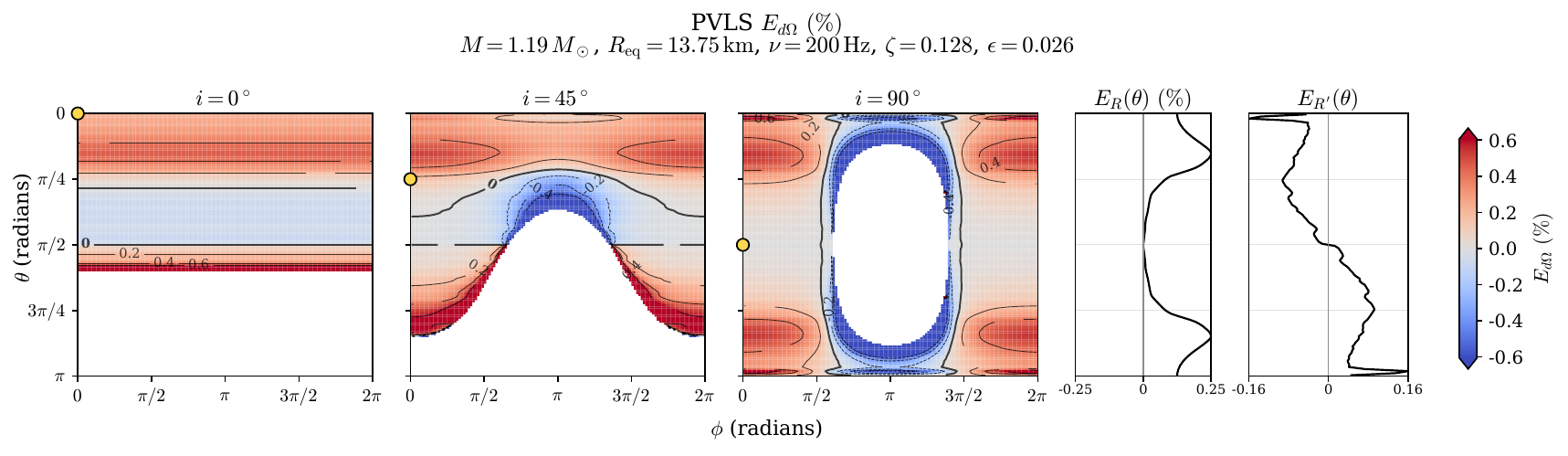}
    \caption{
    Example $E_{d\Omega}$ maps over the neutron star surface for a spin frequency of 200 Hz. These represent a slow-rotation (low $\epsilon$) case constructed with EOS CMF2. For each plot, $\theta$ is the colatitude, $\phi$ is the longitude, and the yellow dot represents the location of the distant observer relative to the angular coordinates of the surface of the star. The first three panels of each row are $E_{d\Omega}$ surface maps at different inclinations; the fourth panel is the error in the radius in percentage relative to the \texttt{RNS} surface; and the fifth is the normalized error in the radius derivative relative to the \texttt{RNS} surface. For each shape function, the color scale is set from the $i=45^\circ$ panel and made symmetric about zero using the larger absolute value of the 5th and 95th percentiles; the same scale is used for all three inclination panels in that row. The white no-data regions are the far side of the neutron star where no photons manage to make it to the observer. The top row is an example of the AM shape function, which for this $\epsilon$ retains an elliptical error profile. The bottom row is an example of PVLS' irregular 'bump' type error. Note that in the PVLS case, $E_{R'}$ does not reflect the derivative of $E_{R}$ because $R'$ for PVLS is computed utilizing a separate neural network. Furthermore, the spike in the error close to $\theta = 0$ or $\theta = \pi$ is due to an issue with the ANN described in Section \ref{sec:E_r'_errors}.}
    \label{fig:DiffSolidAngleExample-200Hz}
\end{figure*}

In each example, as in Figure \ref{fig:DiffSolidAngleExample-200Hz}, we show a horizontal row with 5 panels. The first 3 panels from left to right represent the surface of the star, plotted as a rectangle. The horizontal axis in each case is the longitudinal angle $\phi$ of the infinitesimal emitting area. The observer is located at $\phi = 0$, so that the center of the plot at $\phi=\pi$ is the longitude opposite to the observer. The vertical axes correspond to the colatitude $\theta$ of the emitting area, where $\theta = 0$ is the north spin axis and $\theta=\pi$ is the south spin axis. Each of the three panels shows the percent error in $d\Omega$ from the point of view of an observer located at the inclination angle given in the label above the panel, and indicated with a yellow circle on the vertical axis. The color bar at the far right of the figure shows the percent error in the differential solid angle of a surface element at the given location on the star due to the use of the indicated approximate shape function, as viewed by an observer located at the given inclination angle. The maximum color saturation on either end corresponds to the larger absolute value of the 5th and 95th percentiles.
The white regions are the parts of the star that are not visible to the observer.
The fourth and fifth panels show the profile of the errors in the radius and radius derivative (with respect to $\theta$). Figure~\ref{fig:DiffSolidAngleExample-200Hz} illustrates errors for a neutron star spinning at 200~Hz, typical of the slower RPPs observed by NICER, while Figure~\ref{fig:DiffSolidAngleAMNicer} examines the use of the AM shape function in current NICER pulse profile analysis of the fastest RPP, spinning close to 350~Hz.  Figure~\ref{fig:DiffSolidAngleExample-600Hz} illustrates errors at higher $\epsilon$ where the AM shape function has a bloated error profile, and Figure~\ref{fig:DiffSolidAngleDiags} illustrates the relative weight of different components of the $d\Omega$ error and the error's overall impact at different locations.

\subsubsection{Observer at $i = 0^\circ$}

Observations of the rotating neutron star down the rotation axis are shown in the $i=0^\circ$ panels in each example. This simple case shows the relative contributions of $E_R$ and $E_{R'}$ to $E_{d\Omega}$ at each latitude, and is easiest to interpret since they are independent of longitude on the star. From Table~\ref{tab:coslambda}, when $i=0$, $\cos\lambda = 1$, simplifying equation~\ref{eq:domega_percent_error_analytic}. Light emitted from the north pole ($\theta=0$) travels to the observer with $\alpha = \sigma = 0$ for this case, meaning that the error in the differential solid angle is approximately twice the error in the radius at this latitude.

In Figure \ref{fig:DiffSolidAngleExample-200Hz}, the stellar model is the same as the stellar model shown in the top left panels of Figures \ref{fig:shape_morphology_examples} and \ref{fig:dr_dmu_errors_morphology_examples}. 
In Figure \ref{fig:DiffSolidAngleExample-200Hz}, the value of $E_{R}$ at the north pole is approximately $+0.11\%$ for the AM shape function, and the error $E_{d\Omega}$ shown in the leftmost panel at this latitude is approximately $+0.22\%$. Similarly, the lower panels show that the PVLS errors are about 50\% larger at the pole at $E_{R} = +0.16\%$ and $E_{d\Omega} = +0.32\%$, respectively. At larger values of $\theta$, both $\alpha$ and $\sigma$ increase, leading to contributions from the projection error term given in equation (\ref{eq:projection_error_analytic}). Since $M/R$ is small, the projection term is dominated by the $E_{R'}$ term, which has the same sign as $E_{R'}$ since $\cos\lambda = 1$. In the range of $0 \le \theta \le \pi/4$, the AM radius error decreases, while the radius derivative error is negative with an increasing magnitude. The interplay of these two terms leads to $E_{d\Omega}$ decreasing to close to zero around $\pi/4$ for this example. This should be contrasted with the PVLS shape function, which for this star has an irregular bump error in $E_R$ that rises to close to 0.3\% in this region. This leads to a larger error of approximately 0.4\% in $E_{d\Omega}$ in this region. 

Moving closer to the equator from $\pi/4 \le \theta \le \pi/2$, the AM function's derivative error is negative, leading to negative error in $E_{d\Omega}$. The PVLS shape has much lower radial error closer to the equator, leading to smaller errors in the differential solid angle in this region. At this viewing angle, the limb, $\cos\sigma = 0$, corresponds to the largest value of $\theta$. In the limit of the limb, the dominant term in the error comes from the error in the derivative at the latitude of the limb, which is positive in the southern hemisphere for the AM and PVLS shape functions. The general trends for observers viewing down the spin axis described here can also be seen in the left-most panels of the examples shown in Figures~\ref{fig:DiffSolidAngleAMNicer} and \ref{fig:DiffSolidAngleExample-600Hz}.

\subsubsection{Observer at $i = 90^\circ$}

The case of an observer looking in the direction of the star's equatorial plane is shown in the 3rd panel of each example, labeled $i = 90^\circ$, and depends on the longitude of the emitting region. The photons emitted directly to the observer from $\phi=0$ and $\theta=\pi/2$ have $\alpha=\sigma=0$, so the error in the differential solid angle at this point is simply twice the error in the radius. The three shape functions featured in these examples (AM, SPYY, and PVLS) are constructed to have zero error at the equator, so the error $E_{d\Omega}$ vanishes at this point for all cases and is small in the region near to this point, as can be seen in all of the $i=90^\circ$ panels in Figures \ref{fig:DiffSolidAngleExample-200Hz}, \ref{fig:DiffSolidAngleAMNicer}, and \ref{fig:DiffSolidAngleExample-600Hz}. Due to the geometrical definitions, when the observer is in the equatorial plane, $\cos\lambda = - \cos\theta \cos \phi/\sin\psi$. 
This means that $\cos\lambda$ has the opposite sign in the northern and southern hemispheres, so that the contribution of the radius derivative error to the solid angle error is symmetric through reflections through the equator, and zero on the equator.

In the "front" side of the star (longitudes with $\phi\le\pi/2$ or $\phi \ge 3\pi/2$), $\cos\lambda < 0$ in the northern hemisphere and $\cos\lambda > 0$ in the southern hemisphere (as can be seen from Table~\ref{tab:coslambda} for the special cases of $\phi = 0$ and $\phi=\pi$). At $\phi=\pi/2$ and $\phi=3\pi/2$, $\cos\lambda=0$. 

For this reason, both the radius and radius derivative errors lead to positive solid angle error for the AM shape on the side facing the observer, as shown in Figures \ref{fig:DiffSolidAngleExample-200Hz} and \ref{fig:DiffSolidAngleAMNicer}. This is also true for the PVLS example shown in Figure \ref{fig:DiffSolidAngleExample-200Hz}, although the properties of the derivative error are more complicated. 

The far side of the star, $\pi/2 \le \phi \le 3\pi/2$, has positive $\cos\lambda$ in the north and negative in the south. The limb is confined to the far side of the star for this inclination angle, so the sign of the error near the limb is given by the sign of the derivative error in the northern hemisphere. This results in negative error in the differential solid angle in a circle around the eclipsed region for the examples in Figures \ref{fig:DiffSolidAngleExample-200Hz} and \ref{fig:DiffSolidAngleAMNicer}. While the error patterns for the more rapidly rotating case shown in Figure \ref{fig:DiffSolidAngleExample-600Hz} are more complicated, they can be understood using the same arguments.



\subsubsection{Observer at $i=45^\circ$}

The error patterns for observers with inclination angles at mid-latitudes are more complicated than the cases for $i=0^\circ$ or $90^\circ$. As a representative case, we have chosen to provide error maps on the example stars for an observer at $i=45^\circ$.
The sign of the contributions from the radius derivative error can be understood using the values of $\cos\lambda$ given in Table~\ref{tab:coslambda}. Along longitude $\phi = 0$, the sign of $\cos\lambda$ changes at the colatitude of the observer ($\pi/4$ in this case). Similarly, along longitude $\phi = \pi$, the sign changes at the colatitude antipodal to the observer (at $3\pi/4$ for this example). 

In the region between the north pole and $\theta = \pi/4$, there is very little dependence of the error on longitude, and the errors are very similar to the $i=0^\circ$ panel for the slowly rotating star shown in Figure~\ref{fig:DiffSolidAngleExample-200Hz}. At colatitudes in the range from $\pi/4 < \theta < 3\pi/4$, $\cos\lambda$ is positive for all longitudes. Since the radius derivative errors are negative in the north and positive in the south, this leads to the errors in the solid angle having the same sign in the region close to the eclipse. The star used in Figure~\ref{fig:DiffSolidAngleExample-200Hz} does not have very strong gravitational lensing, so the region near the star's south pole is not visible. It is useful to contrast this with the situation shown in Figure~\ref{fig:DiffSolidAngleAMNicer}, which has a stronger gravitational field and a smaller eclipsed region. For the star in Figure~\ref{fig:DiffSolidAngleAMNicer}, the region near the south pole with $3\pi/4 < \theta < \pi$ is visible. For this range of latitudes, $\cos\lambda$ is negative near $\phi = \pi$, leading to a large negative contribution to the solid angle error for light emitted near the star's south pole.

\subsubsection{Errors in a NICER-like Rotation Powered Pulsar}

\begin{figure*}[!ht]
    \centering    \includegraphics[width=\linewidth]{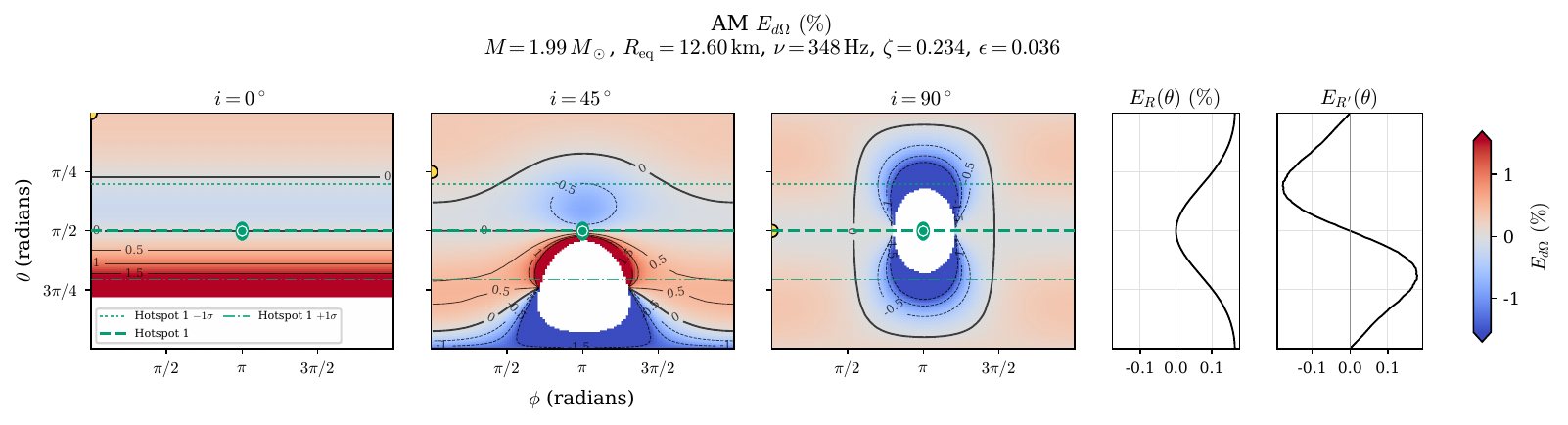}
    \includegraphics[width=\linewidth]{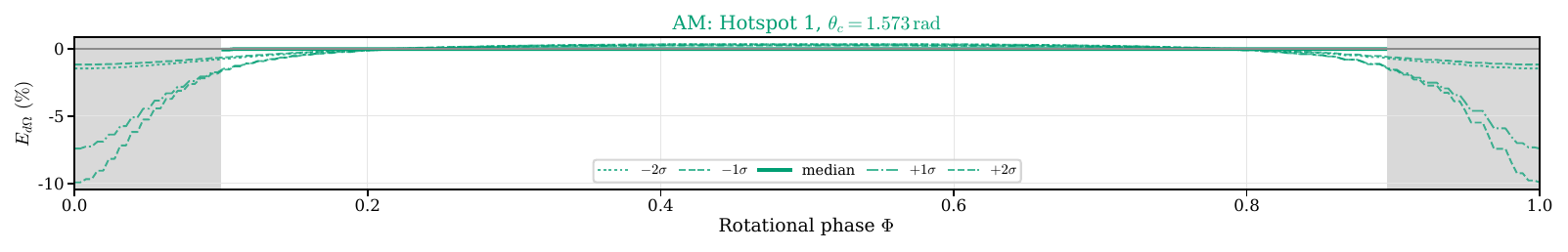}
    \caption{Top: Example $E_{d\Omega}$ maps over the neutron star surface for a star with parameters similar to those inferred from NICER observations of PSR~J0740+6620 \cite{2024ApJ...974..295Dittmann,2024ApJ...974..294Salmi}. The neutron star was computed using EOS HLPS2, and the AM shape function was used to approximate the shape.
    The plots are constructed similarly to Figure \ref{fig:DiffSolidAngleExample-200Hz}, but the location of one of the inferred hotspots from \cite{2024ApJ...974..295Dittmann} and its travel path as the star rotates are overlaid in green, along with the $\pm 1 \sigma$ limits on the center of the spots. The other hotspot is located at a very similar latitude and has similar uncertainties, but with a longitudinal offset, so only one is displayed for simplicity.
    Bottom: The error in the differential solid angle of the hotspot as a function of the rotational phase, using the $i=90^\circ$ model. Curves of $E_{d\Omega}$ are shown for infinitesimal spots located at the median, $\pm 1 \sigma$, and $\pm 2 \sigma$ values for the spot's location.
     The errors reflect the hotspots traveling horizontally across the $i=90^\circ$ error map due to the spin of the neutron star, and can be considerable near the limb. The gray zones indicate regions where the median position of the hotspot is eclipsed at the back of the neutron star.}
    \label{fig:DiffSolidAngleAMNicer}
\end{figure*}

The parameter estimation codes used to analyze the NICER observations of the RPPs make use of the AM shape function. Since the errors in the AM shape grow with spin frequency, it is useful to consider the most rapidly rotating RPP, PSR~J0740+6620, in order to evaluate the worst-case scenario. We constructed a star using EOS HLPS2 with $M=1.99 M_\odot$, $R_{\rm eq} = 12.6$ km, and $\nu = 348$ Hz, values similar to the median values found in the analyses of NICER data \cite{2024ApJ...974..295Dittmann,2024ApJ...974..294Salmi}. The AM shape function and its derivative shown in Figure~\ref{fig:DiffSolidAngleAMNicer} have errors that are $\le 0.2\%$ over the star's surface. However, the regions near the limbs have $E_{d\Omega}$ errors that are much larger. The observer's inclination angle is known to be very close to $i = 90^\circ$ from radio observations \cite{OBS_MASS_SHAPIRO_Cromartie_2019,EX_NS_LARGE_2021ApJ...915L..12F}, so the third panel for the example in Figure~\ref{fig:DiffSolidAngleAMNicer} is the one that best represents this pulsar. The median locations and angular widths of the first of two spots inferred in reference \cite{2024ApJ...974..295Dittmann} are shown as a green circle, along with horizontal lines showing the $\pm 1 \sigma$ locations of the spot center. Due to the similarity in latitude and uncertainties of the two hotspots, only one is displayed here for simplicity. The two spots inferred in reference \cite{2024ApJ...974..294Salmi} are slightly offset from these values, but their $1 \sigma$ range of locations is very similar.

We further show the differential solid angle errors as a function of pulsar phase for one of the hotspots modeled by \cite{2024ApJ...974..295Dittmann}
in the lower panel of Figure~\ref{fig:DiffSolidAngleAMNicer}. As would be expected, the errors for an infinitesimal spot located on the equator are minimal. However, spots located at the $\pm 1 \sigma$ colatitudes travel through the region with the largest errors, which leads to the computed spots underestimating the differential solid angle by roughly 10\% for about 20\% of the star's rotational phase. The $\pm 2 \sigma$ colatitudes (not shown in the upper panels) are much closer to the poles, where the errors are much smaller.

Whether this amount of error seriously affects the accuracy of the equatorial radius estimate is a more complicated problem. The error estimates that we show here are for an infinitesimal emitting area. However, realistic hotspots have larger angular extents, which tend to smooth out the effect of errors near the eclipses \cite{2015ApJ...811..144Baubock}.
The predictions about the accuracy of the radius estimate from a Bayesian analysis require assumptions about the relative number of counts \cite{2013ApJ...776...19Lo} due to the spot and the background. Earlier work \cite{2015ApJ...808...31Miller} compared the errors in equatorial radius resulting if a perfectly spherical stellar surface was assumed in the analysis of data for a star spinning at 600 Hz, with an extended hotspot, viewed at an inclination angle of $90^\circ$, a case that would have much larger errors than what we show here. Their results (see Figure 2d of \cite{2015ApJ...808...31Miller}) show a small bias in the estimated radius. However, it is also known that an incorrect estimate of the unpulsed background \cite{2021ApJ...918L..28M} can lead to a much larger bias in the inferred radius. 

In the flux computation, the differential solid angle is multiplied by the specific intensity of light emitted from the area. The atmospheres of the RPPs are usually modeled with a single-element atmosphere, such as Hydrogen \cite{OBS_HEINKE_QLMXB_2006ApJ...644.1090H}, which is highly beamed towards the normal. Emission from the limbs is then highly suppressed since the light is emitted nearly tangent to the surface. The situation for the AMXPs is more complicated, since Compton scattering of photons can lead to anti-beaming \cite{2023A&A...678A..99Bobrikova}.

Although it is not clear whether the errors caused by the shape function cause significant errors in the radius estimate, it makes sense to construct an improved shape function for future analysis of timing data.

\subsubsection{Rapid Rotation}

Some of the accreting ms-period X-ray pulsars and X-ray bursters spin more rapidly, with spin frequencies ranging from 400 - 600 Hz, as listed in Table~\ref{tab:obs_ns_examples}. The present versions of the codes used to analyze pulse profiles make use of the AM shape. The AM shape function was tuned for use with the more slowly rotating rotation-powered pulsars with computations restricted to $\epsilon \le 0.1$. Neutron stars spinning with frequencies close to 600 Hz will have larger values of $\epsilon$, so the AM shape will likely be inadequate for the analysis of more rapidly rotating stars.

\begin{figure*}[!ht]
    \centering
    \includegraphics[width=0.9\linewidth]{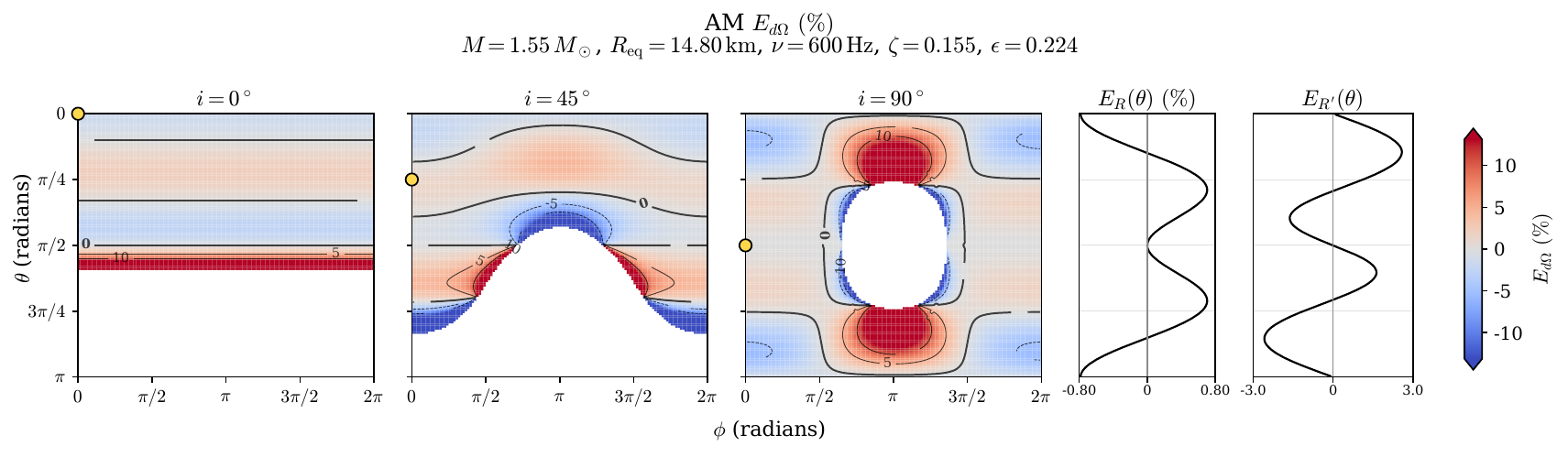}
    \includegraphics[width=0.9\linewidth]{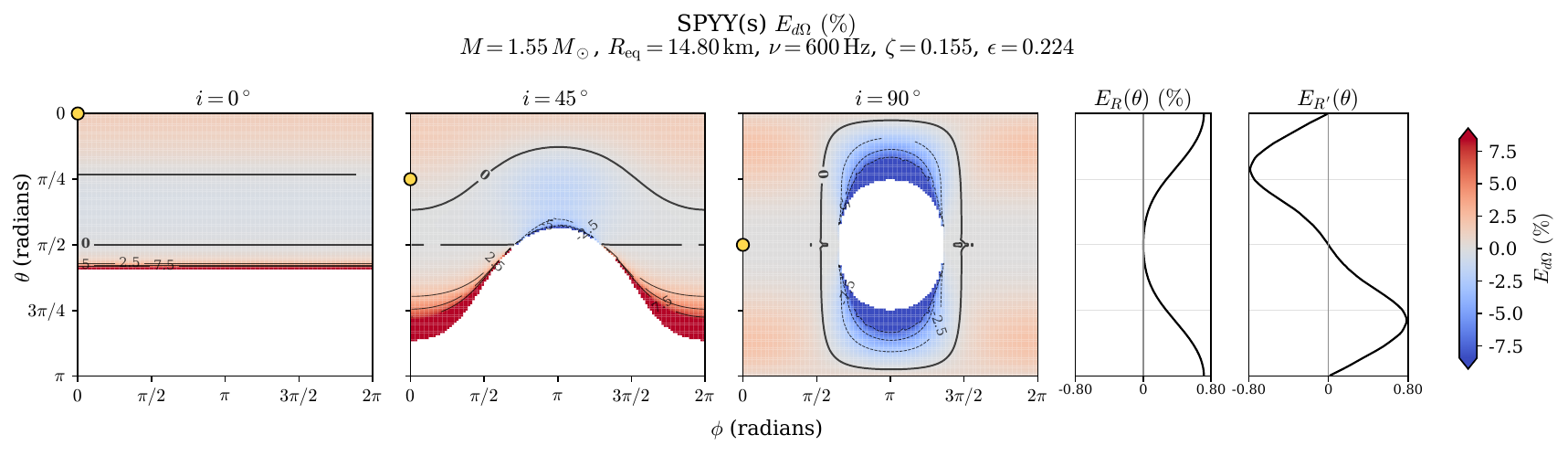}
    \includegraphics[width=0.9\linewidth]{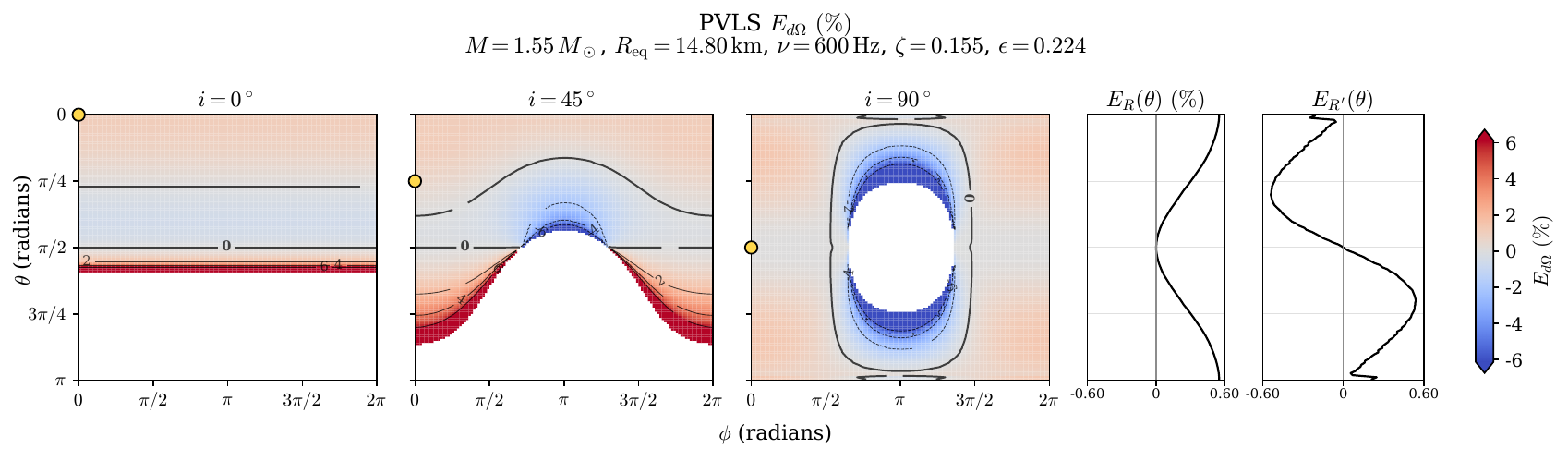}
    \caption{
    Example $E_{d\Omega}$ maps over the neutron star surface. These represent fast-rotation (higher $\epsilon$) cases with EOS CMF2. The plots are constructed similarly to Figure \ref{fig:DiffSolidAngleExample-200Hz}. Notice that AM, now with a bloated mid-colatitude error, has a much larger $E_{R'}$ and correspondingly larger $E_{d\Omega}$ than expected from the $2E_{R}$ area contribution alone. This may be contrasted with SPYY(s) with the same parameters, whose maximum $E_{R}(\theta)$ is comparable in scale to AM but has a smaller maximal $E_{R'}(\theta)$ and a correspondingly smaller scale of $E_{d\Omega}$. PVLS is extremely similar to SPYY(s) in this example, but has a small spike in $E_{R'}$ towards the poles due to the issue described in Section \ref{sec:E_r'_errors}.}
    \label{fig:DiffSolidAngleExample-600Hz}
\end{figure*}

In Figure~\ref{fig:DiffSolidAngleExample-600Hz}, we show the $E_{d\Omega}$ error map for one neutron star rotating at a frequency of 600 Hz and $\epsilon = 0.224$. All three of the shape functions shown (AM, SPYY(s), and PVLS) have $|E_R(\theta)| < 1\%$ everywhere on the star. In the cases of SPYY(s) and PVLS, the shape function training data included values of $\epsilon$ as large as this star's, and the maximum value of $E_{R'}$ is similar to the maximum value of $E_R$. 
This should be contrasted with the AM shape function, whose construction did not include such large values of $\epsilon$ and whose maximum $E_{R'}$ values are about 4 times larger than its $E_R$ maximum. The result for the AM shape is an error function with more oscillations in sign than the SPYY(s) and PVLS functions, and larger errors in the $E_{d\Omega}$ maps. The AM map for $i=90^\circ$ shows two large lobes of large positive errors on the side of the star opposite to the observer. The SPYY(s) and PVLS functions have similar error patterns and magnitudes, where most of the errors are concentrated along the limbs.

Due to the large errors in solid angle shown in the AM error maps in Figure~\ref{fig:DiffSolidAngleExample-600Hz}, we recommend that either the SPYY (with the appropriate coefficient set depending on $\epsilon$), PVLS, or an improved shape function designed for rapid rotation be used for the AMXPs or other rapid rotators.

\subsubsection{Weighting $E_{d\Omega}$}

$E_{d\Omega}$ is computed at a point on the surface of the neutron star, but its contribution to the integrated projected solid angle $\Omega_{\rm proj}$ or flux $F_{E}$ is weighted by $d\Omega$ at that point. For example, near the limb, the local $d\Omega$ at each point is minuscule, but $E_{d\Omega}$ is large. This raises the question of whether the error near the limb is an insignificant contribution to the overall error, the dominant effect, or roughly equivalent to the error contributed by the rest of the surface. 

\begin{figure*}[!ht]
    \centering
    \includegraphics[width=0.9\linewidth]{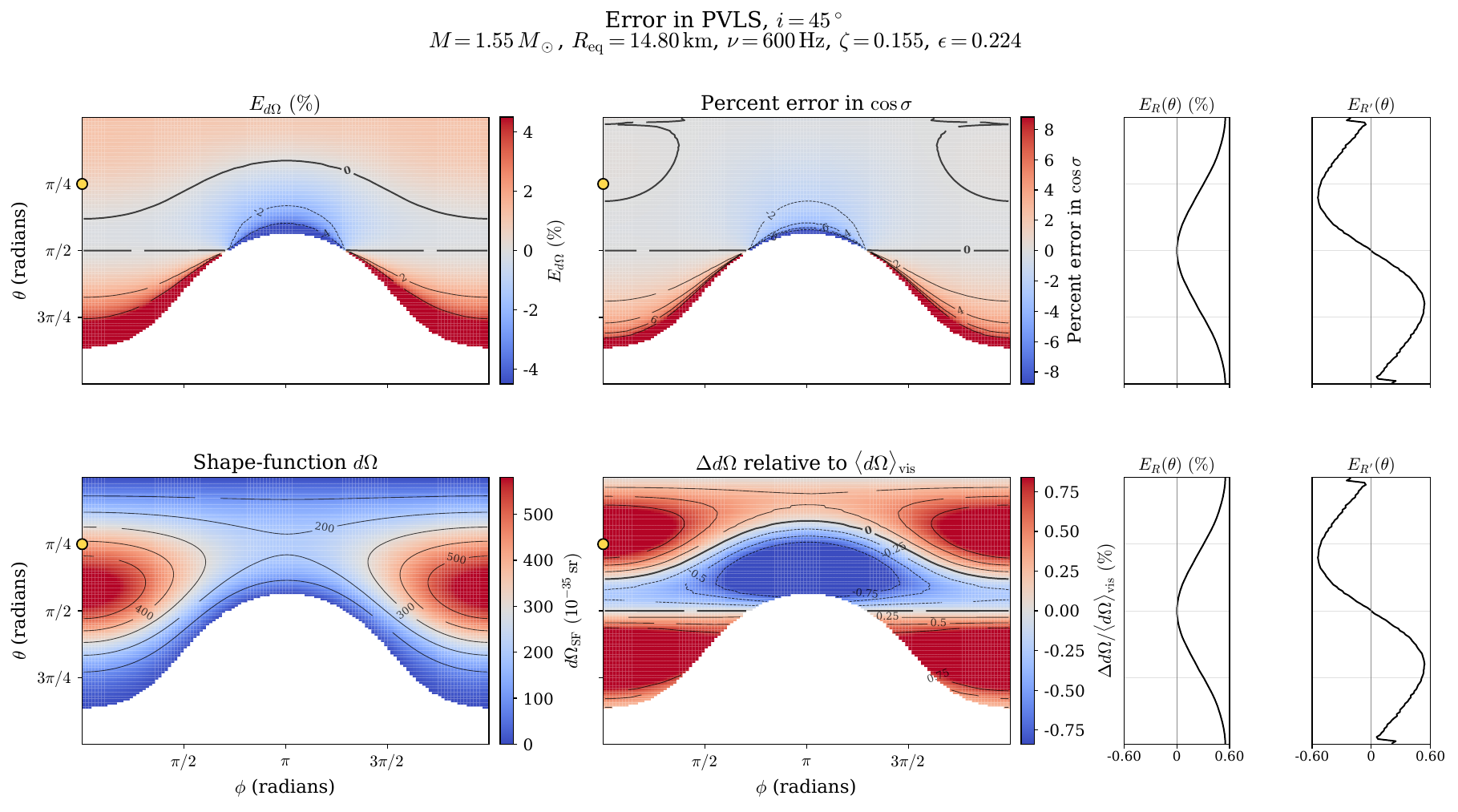}
    \caption{
    Breakdown of $E_{d\Omega}$ for one PVLS example relative to an \texttt{RNS} reference surface computed using the CMF2 EOS. For the three error maps, the color scale is symmetric about zero, with its limits set by the larger absolute value of the 5th and 95th percentiles of the data. Top left: percent error in $d\Omega$ over the visible surface. Top right: percent error in the projection factor $\cos\sigma$, showing where surface-normal errors contribute to the local solid angle error. Bottom left: the shape function value of $d\Omega$, which shows how strongly different visible patches contribute to the integrated solid angle. Bottom middle: $\Delta(d\Omega)$ normalized by the visible-surface mean $\langle d\Omega_{\rm RNS}\rangle_{\rm vis}$, showing the local error relative to a typical visible patch. The right-hand line panels show the corresponding colatitude profiles of $E_R(\theta)$ and $E_{R'}(\theta)$, illustrating how radius and slope errors are reflected in the differential solid angle maps. \new{The $E_R$ and $E_{R'}$ plots on the top row are repeated on the bottom row for ease of comparison with the surface map.}}    
    \label{fig:DiffSolidAngleDiags}
\end{figure*}

Figure~\ref{fig:DiffSolidAngleDiags} examines this for a PVLS $i=45^\circ$ case in more detail. The top-left panel repeats the percent error in $d\Omega$ shown in Figure~\ref{fig:DiffSolidAngleExample-600Hz}. The top-right panel shows the percent error in $\cos\sigma$. These two panels illustrate that the large errors in $d\Omega$ seen near the limbs are mainly attributable to errors in $\cos\sigma$. Farther away from the limbs, the error is close to the sum of $2E_{R}$ and the error in $\cos\sigma$.

The bottom left panel plots $d\Omega$ at each point on the star. It is visibly skewed away from the limb, yet from prior plots we observe that the percent error relative to each point is most concentrated at the limb. 

The bottom middle figure illustrates the local difference normalized by the average visible surface element,
\begin{equation}
\frac{\Delta(d\Omega)}
{\langle d\Omega_{\rm RNS}\rangle_{\rm vis}}
=
\frac{
d\Omega_{\rm sf}-d\Omega_{\rm RNS}
}{
\langle d\Omega_{\rm RNS}\rangle_{\rm vis}
} ,
\label{eq:domega_visible_mean_difference}
\end{equation}
where $\langle d\Omega_{\rm RNS}\rangle_{\rm vis}$ is the mean \texttt{RNS} differential solid angle over the visible surface. This plot reveals that the contribution to the total error in the solid angle is about as large from regions close to the limb as from those far from the limb. Nowhere on the surface can the error added to the overall solid angle be discounted.

As an aside, the integrated projected solid angle $\Omega_{\rm proj} = \int d\Omega$ does not provide a useful quantification of the error.  Using it, we run into two issues. First, the ultimate observable is the flux
\begin{equation}
F_E = \int_{\rm vis} \Upsilon(\theta,\phi,\ldots)\,d\Omega.
\label{eq:flux_weighted_domega_integral}
\end{equation}
The intensity may vary by location on the surface, such as with hotspots or latitude-dependent temperature changes. The dependence of the intensity on the emission angle with respect to the local normal also introduces dependencies on position. These effects must be multiplied by the differential solid angle at each location to obtain the correct flux. The total integrated solid angle is only meaningful in the context of uniform and isotropic emission intensity.
Second, the integrated solid angle may be deceptively accurate even when the underlying differential solid angle errors are not accurate,  due to errors with opposite signs canceling. These canceling errors may be visualized by summing up positive and negative error contributions seen in the bottom middle panel of Figure~\ref{fig:DiffSolidAngleDiags}.
As such, we have omitted any further analysis of this quantity.

\section{Conclusions}
\label{sec:conclusions}

\subsection{Main findings}

We compared five shape functions against surfaces computed with \texttt{RNS} over a broad range of compactness and spin parameters. Despite all being built on the assumption of quasi-universality, the shape functions differ from each other and from the \texttt{RNS} surfaces. The magnitude, dependence on colatitude, scaling with compactness, and scaling with spin of their errors depend on the form of the shape function. MLCB, SPYY, and PVLS at moderate and large $\epsilon$ tend to produce smooth errors corresponding to surfaces that are more or less elliptical compared to the \texttt{RNS} surface, while BBPO and AM develop mid-colatitude errors at sufficiently rapid rotation. PVLS additionally develops irregular errors at small $\epsilon$, as its training data omits part of this range. The corresponding errors in $R'(\theta)$ broadly follow these same morphologies, although PVLS is unusual because its derivative is predicted by a separate neural network.

Despite these errors, we find support for the assumption of quasi-universality. For each shape function, the extremal radius errors from different EOSs form a structured surface in compactness-spin parameter space without visible dependence on the choice of EOS. A simple correction in colatitude, compactness, and spin reduces the errors in AM and SPYY(s) by approximately an order of magnitude. After correction, the bulk of the remaining errors between the 5th and 95th percentiles lies within approximately $\pm 0.5\%$ for AM, SPYY(s), and PVLS. Leaving each EOS out of the correction fit in turn and evaluating the errors on that excluded EOS produces very similar results. Some EOS-dependent offsets from zero remain, particularly for PVLS, so our results do not establish exact universality. These residual offsets may represent either limitations of the particular shape functions or a potential limit to the quasi-universality assumption.

We analyzed the treatment of the polar radius in the PVLS shape function \cite{SF_Papigkiotis_ML}. When the \texttt{RNS}-computed polar radius is supplied to the PVLS model, as in the original analysis, the ANN reproduces the remaining surface extremely accurately. However, the polar radius is not known independently in applications such as pulse-profile modeling and must itself be predicted from $M$, $R_{\rm eq}$, and the spin. Using the PVLS analytical fit for $R_{\rm pole}$ increases the maximum relative surface error over their dataset from $0.25\%$ to $2.77\%$. The error in the predicted polar radius dominates over the error in the ANN reconstruction of the surface between the pole and equator. This indicates that accurately predicting the polar radius, or equivalently the eccentricity, is one of the central remaining problems in constructing more accurate shape functions.

We also explored the relationship between the radial error and the differential solid angle error within the context of the oblate Schwarzschild approximation. Within the regime where the Beloborodov approximation is accurate, the self-lensing and Jacobian terms cancel. The remaining error separates into an area contribution, approximately $2E_R$, and a projection contribution. The projection contribution depends primarily on $E_{R'}$, since an error in the radial derivative changes the direction of the local surface normal. For photons emitted close to the radial direction, the projection contribution is small and $E_{d\Omega}\simeq2E_R$. Towards the limb, the projection contribution is amplified as $\cos\sigma$ approaches zero. The resulting $E_{d\Omega}$ depends on where that error occurs on the surface and on the viewing geometry.

This behavior is important even for neutron stars rotating at the frequencies presently observed by NICER. For our model with parameters similar to PSR~J0740+6620, the AM errors in $R(\theta)$ and $R'(\theta)$ remain below approximately $0.2\%$, yet the corresponding errors in $d\Omega$ become much larger near the limb. Hotspots placed at the $\pm1\sigma$ colatitudes inferred from NICER observations pass through these regions and can have their differential solid angle underestimated by approximately $10\%$ over about $20\%$ of the rotational phase. 
This does not by itself establish a comparable bias in the inferred radius, but it demonstrates that the shape of the flux may be altered and that sub-percent errors in the stellar surface may nonetheless result in \new{much larger errors in the geometric contribution to the flux from individual surface elements, potentially altering the pulse profile.} 
Our 600 Hz example further illustrates these effects: AM and SPYY(s) can have radius errors of comparable magnitude while differing substantially in $R'(\theta)$ and consequently in $E_{d\Omega}$.

Finally, the large relative errors near the limb are not minimized by $d\Omega$ becoming small there. In the example examined here, the amount of the local error near the limb is comparable to that contributed by other parts of the visible surface when normalized to a typical visible surface element. At the same time, integrating $d\Omega$ over the entire visible surface can hide these errors through cancellation between positive and negative contributions. Since the emitted intensity need not be uniform over the surface, such cancellation need not persist in the observed flux. The integrated projected solid angle is a poor measure of shape function accuracy. The relevant errors must instead be considered locally through both $R(\theta)$ and $R'(\theta)$ and through their propagation into $d\Omega$.

\subsection{Limitations}

These conclusions apply to the five shape functions examined here: MLCB, BBPO, AM, SPYY, and PVLS. For SPYY, the slow and fast coefficient sets were treated separately. They also apply to the EOS sample described in Sec.~\ref{subsec:eos_sample}. This sample includes a controlled soft-to-stiff sequence in the form of the HLPS family and several additional modern EOSs outside most of the calibration sets used by the shape functions, but it is not an exhaustive EOS survey. By the assumption of quasi-universality, and by our choice of a diverse set of EOSs spanning a realistic range of stiffness, we expect these results will generalize well. Nonetheless, a different EOS set could reveal behavior not captured here, especially if it occupies a poorly sampled region of compactness-spin space or manages to produce surface shapes outside the range covered by the present EOSs as run through \texttt{RNS}.

The comparison was carried out over the parameter space described above. For differential solid angle analysis, we are further restricted in the compactness to $\zeta\leq 1/3.52$. Above this compactness, Schwarzschild light bending can produce multiple images of the same surface region, which is not modeled by our NS-SWORD implementation.

The \texttt{RNS} reference surfaces were generated for a range of central densities, spins, and EOSs. The masses and radii emerged from this computation, and then they were compared to the literature shape functions at the same model parameters. The results therefore depend on the accuracy of the \texttt{RNS} surfaces, the interpolation onto the colatitude grid, and the density of the model sampling. 

The propagation into $d\Omega$ was calculated within the oblate Schwarzschild approximation. This lets us isolate the effect of replacing the \texttt{RNS} reference model for a given EOS with a shape function while computing the overall result in Schwarzschild spacetime. It does not include the full spacetime of a rotating neutron star, including frame dragging or the rotational quadrupole of the exterior metric.

The analytic cancellation between the self-lensing and Jacobian terms uses the Beloborodov approximation. The cancellation is useful only where the local compactness is sufficiently low (approximately $M/R(\theta)\leq 0.25$).

This work focuses on modeling the geometry. We compare errors in $R(\theta)$, $R'(\theta)$, and $d\Omega$. We do not generate spectra, calculate the full flux, or carry out parameter inference for an inferred radius in this work. Our results show where shape function errors affect the flux calculation. The size of the final observational bias will further depend on the temperature distribution, atmosphere model, viewing inclination, and details of the observatory which makes the observations.

\subsection{Recommendations for Developing Better Shape Functions}

Our results give us a basis to recommend future directions for a better shape function. We suggest the following requirements:

\begin{enumerate}

\item \textbf{Known quantities should be enforced within the functional form where possible.} Any shape function should enforce known quantities, such as the equatorial radius, ideally within its functional form. As a counterexample, the MLCB shape function does not do this, and thus has considerable errors even at the equator. Such enforced quantities help to constrain the shape function and improve its accuracy.

\item \textbf{The fitted parameter space should be well-defined and respected.} Any better shape function must have a well-defined parameter space and have a large dataset of neutron stars computed with a variety of EOSs across that parameter space. \new{An advantage of PVLS is that it was trained on 70 EOSs and approximately $4\times10^4$ rotating neutron star models, compared to the much smaller EOS collection available for the earlier MLCB shape function. The shape function should then be considered for use only within those confines. } 
Some of the largest errors found in this work occur when shape functions are evaluated outside the parameter ranges used in their construction. This includes AM at large $\epsilon$, SPYY(s) beyond its slow-rotation range, and PVLS at very small $\epsilon$ and $\zeta$.

There may also be value in constructing shape functions for particular applications rather than attempting to cover every possible compact star with one fit. SPYY was built with this intuition in mind, although this introduces difficulties in choosing the coefficient set appropriate for the context. \new{Another example of this approach is a recent shape function published \cite{2026arXiv260900172M} after this paper was submitted, which was constructed specifically for the analysis of PSR~J1614-2230.} 
A shape function could, for example, be fitted densely over a restricted range of compactness and spin relevant to a particular class of observed neutron stars, such as low-compactness, low-mass neutron stars. Bare quark stars may similarly require a separate treatment from hadronic neutron stars, since their surface shapes need not follow the same quasi-universal relation.

\item \textbf{The polar radius or eccentricity should be modeled accurately.} Modeling the eccentricity or the polar radius of rapidly rotating neutron stars accurately may be one of the key problems affecting the accuracy of shape functions. This is implied by the results of the machine learning shape function PVLS. It has a highly accurate surface fit given knowledge of the equatorial and polar radii. However, it is considerably less accurate if forced to model the polar radius using its polynomial fit. In our implementation, replacing the \texttt{RNS} reference polar radius that would not ordinarily be known ahead of time with this fit increases the maximum relative surface error from approximately $0.25\%$ to $2.77\%$. This suggests that improving the modeling of $R_{\rm pole}$ is a method to reduce the relative error to as low as $0.25\%$.

\item \textbf{Errors in both $R(\theta)$ and $R'(\theta)$ should be minimized, and should vary smoothly with colatitude.} A better shape function should keep $E_{R}(\theta)$ small over the colatitude-compactness-spin range relevant to observed neutron stars. The area term alone gives an approximate contribution $2E_{R}$ to $E_{d\Omega}$, even before the projection term is considered. The requirements depend on the required accuracy of $d\Omega$ in the flux model.

The shape function should also keep errors in $R'(\theta)$ small. The differential solid angle depends on the projected area of each surface element. The projection term depends on $\sigma$, the angle between the emitted ray and the surface normal. Since the surface normal depends on $R'(\theta)$, a shape function with a small radius error can still produce a larger local error in $d\Omega$, particularly towards the limb, if its derivative is inaccurate.

The error in $R(\theta)$ should vary smoothly with colatitude. A smooth one-sided error, such as a surface that is more or less elliptical than required, can be easily understood as an error in the coefficients of the shape function and potentially accounted for in corrections to analyses that utilized it. Sharp bumps, valleys, or irregular errors are less desirable because they are more difficult to account for in future corrections and can produce correspondingly complicated errors in $R'(\theta)$.

\item \textbf{Quasi-universality should be tested across EOSs.} The errors should not appear to depend on the choice of EOS after the colatitude, spin parameter, and compactness allowed by the EOS are accounted for. Our correction analysis supports quasi-universality over the parameter space examined here: after correction, the bulk of the remaining errors between the 5th and 95th percentiles lies within approximately $\pm0.5\%$. Leaving individual EOSs out of the correction fit and evaluating the errors on the excluded EOS produces similar results. However, some EOS-dependent offsets remain, and a small number of rapidly rotating models have errors approaching $\pm4\%$. We cannot determine from this work whether these residuals represent a limit to quasi-universality or limitations of the particular fitting forms. Future shape functions should be tested for their ability to generalize to EOSs not included in their calibration set.

\end{enumerate}

Beyond these requirements, the functional form should remain simple enough that its failures can be accounted for. A highly flexible fit, such as those created with machine learning, may reduce the radius error, but if the remaining error varies irregularly over colatitude, then it becomes harder to know how the error will affect $R'(\theta)$, $d\Omega$, and the flux. That is, we wish for any errors to be small, smooth, and interpretable.

It may also be fruitful to fit a correction to an existing good shape function, or recompute its coefficients, rather than constructing a new shape function from scratch. For example, the SPYY shape function \cite{SF_2021_Silva} might be improved in range and accuracy by utilizing the polar radius or eccentricity polynomial from PVLS \cite{SF_Papigkiotis_ML}. Alternatively, older shape functions might remain relevant as new calibration data become available by adding such correction surfaces. Shape functions are primarily empirical fits to numerically computed surfaces at this point, but a well-designed functional form may also align with and provide insight into the underlying physics of neutron star shapes.

In this work, we have investigated the role of the shape function in the solid angle part of the flux calculation and shown how small errors in the stellar surface can produce considerably larger local errors in $d\Omega$. We have not attempted to construct a new shape function. Developing new shape functions appropriate to the parameter ranges and accuracy requirements of particular modeling applications is a natural topic for future work.

\begin{acknowledgments}
We thank Craig Heinke, Gregory Sivakoff, and Shafayat Shawqi for many useful discussions about this research. We thank the referee for a careful reading and useful comments on this work. 
This work was supported by NSERC Discovery Grants RGPIN-2019-06077 and RGPIN-2026-05707 awarded to SMM. JMN was also the recipient of NSERC CGSM and Alberta Graduate Excellence scholarships.
\end{acknowledgments}

\appendix*

\section{Notation Translation for OS geometry}
\label{app:notation_translation}

Several symbols used in the oblate Schwarzschild geometry have not been used consistently across the literature. We follow the angular notation used in the NICER code-verification paper \citep{OS_2019_NICER} where possible, rather than the older notation of \citet{OS_SF_2007_Morsink}. The main exception is the symbol \(\zeta\): in this paper \(\zeta\) denotes the compactness parameter used in the shape function fits, while \citet{OS_2019_NICER} use \(\zeta\) for the observer colatitude. We keep \(i\) for the observer inclination and write the local compactness as \(M/R(\theta)\).

\begin{table*}[!tb]
\centering
\scriptsize
\caption{Translation between the OS-geometry notation used in this paper and two common conventions in the literature. A blank indicates that the cited work does not introduce a symbol for that quantity.}
\label{tab:notation_translation}
\begin{ruledtabular}
\begin{tabular}{llll}
Quantity & This paper & \citet{OS_2019_NICER} & \citet{OS_SF_2007_Morsink} \\
\hline
Spin frequency & $\nu$ & $\nu$ & $P^{-1}$ \\
Angular spin frequency & $\Omega=2\pi\nu$ & $\Omega=2\pi\nu$ & $\Omega=2\pi/P$ \\
Observer inclination & $i$ & $\zeta$ & $i$ \\
Surface colatitude & $\theta$ & $\theta$ & $\theta$ \\
Surface longitude or phase & $\phi$ & $\phi$ & $\phi$ \\
Observer distance & $D$ & $D$ & $D$ \\
Radial direction & $\hat{r}$ & -- & $\mathbf{r}$ \\
Surface normal & $\hat{n}$ & $\mathbf{n}$ & $\mathbf{n}$ \\
Initial photon direction & $\hat{k}_0$ & -- & $\mathbf{l}$ \\
Observer photon direction & $\hat{k}$ & -- & $\mathbf{k}$ \\
Radial--photon angle & $\alpha$ & $\alpha$ & $\alpha$ \\
Bending angle & $\psi$ & $\psi$ & $\psi$ \\
Normal--photon angle & $\sigma$ & $\sigma$ & $\beta$ \\
Radial--normal tilt & $\tau$ & $\tau$ & $\gamma$ \\
Local azimuthal angle & $\lambda$ & $\lambda$ & $\delta$ \\
Surface-slope factor & $q(\theta)$ & $f(\theta)$ & $f(\theta)$ \\
Combined projection factor & $\mathcal{P}$ & -- & -- \\
Compactness & $\zeta=M/R_{\rm eq}$ & -- & $\zeta=GM/(R_{\rm eq}c^2)$ \\
Local compactness &
$M/R(\theta)$ &
\begin{tabular}[c]{@{}l@{}}
$R_{\rm S}/R$,\\
$M/R$
\end{tabular} &
$M/R(\theta)$ \\
Dimensionless spin &
$\epsilon=\Omega^2R_{\rm eq}^3/M$ &
$\bar{\Omega}^2$ &
$\epsilon=\Omega^2R_{\rm eq}^3/M$ \\
Lorentz factor & $\Gamma$ & $\gamma$ & -- \\
Doppler factor & $\mathcal{D}$ & $\delta$ & $\eta$ \\
\end{tabular}
\end{ruledtabular}
\end{table*}

\bibliography{00_citations}

\end{document}